\documentclass[fleqn,onecolumn,superscriptaddress,aps]{revtex4}
\usepackage{epsfig}
\usepackage{verbatim}
\usepackage{array}

\newcommand{\be}{\begin{equation}}
\newcommand{\ee}{\end{equation}}
\newcommand{\ba}{\begin{eqnarray}}
\newcommand{\ea}{\end{eqnarray}}
\newcommand{\non}{\nonumber}

\usepackage{graphicx}
\usepackage{subfigure}
\usepackage{color} 
\usepackage{amsfonts}
\usepackage{amsmath}

\begin{document}
\title{Shortcuts to adiabaticity using flow fields}
\author{Ayoti Patra}
\affiliation{Department of Physics, University of Maryland, College Park, Maryland 20742, USA}
\author{Christopher Jarzynski}
\affiliation{Department of Physics, University of Maryland, College Park, Maryland 20742, USA}
\affiliation{Department of Chemistry and Biochemistry, University of Maryland, College Park, Maryland 20742, USA}
\affiliation{Institute for Physical Science and Technology, University of Maryland, College Park, Maryland 20742, USA}

\begin{abstract}
A {\it shortcut to adiabaticity} is a recipe for generating adiabatic evolution at an arbitrary pace.
Shortcuts have been developed for quantum, classical and (most recently) stochastic dynamics.
A shortcut might involve a {\it counterdiabatic} Hamiltonian that causes a system to follow the adiabatic evolution at all times, or it might utilize a {\it fast-forward} potential, which returns the system to the adiabatic path at the end of the process. 
We develop a general framework for constructing shortcuts to adiabaticity from flow fields that describe the desired adiabatic evolution.
Our approach encompasses quantum, classical and stochastic dynamics, and provides surprisingly compact expressions for both counterdiabatic Hamiltonians and fast-forward potentials.
We illustrate our method with numerical simulations of a model system, and we compare our shortcuts with previously obtained results.
We also consider the semiclassical connections between our quantum and classical shortcuts.
Our method, like the fast-forward approach developed by previous authors, is susceptible to singularities when applied to excited states of quantum systems;
we propose a simple, intuitive criterion for determining whether these singularities will arise, for a given excited state.
\end{abstract}
\maketitle

\section{Introduction}
The acceleration of quantum adiabatic dynamics is a goal shared by a variety of physical applications, including adiabatic quantum computing \cite{Sarandy11,Santos15, Nielsen}, cold atom transport \cite{Chen10,Torrontegui11,Bowler12}, many-body state engineering \cite{dCampo11,dCampo12b,Masuda14,Mukherjee16} and quantum thermodynamics \cite{Salamon09, Rezek09,Skrzypczyk14}. According to the quantum adiabatic theorem \cite{Born28,Griffiths04}, a system evolving under a Hamiltonian $\hat{H}_0(t)$ remains in an instantaneous energy eigenstate if $\hat{H}_0$ changes adiabatically (infinitely slowly). If $\hat{H}_0$ changes rapidly the system develops undesirable excitations, evolving into a superposition of the energy eigenstates. {\it Shortcuts to adiabaticity}~\cite{Torrontegui13} are  protocols that suppress these excitations by means of an auxiliary Hamiltonian $\hat{H}_1(t)$, which is added to $\hat{H}_0(t)$ to achieve the desired evolution. A variety of shortcut protocols have been developed including invariant-based inverse engineering \cite{Chen11,Torrontegui14}, transitionless counterdiabatic driving \cite{Rice03,Rice05,Berry09, Jarzynski13}, fast-forward methods \cite{Masuda10,Torrontegui12,Martinez-Garaot16} and methods based on unitary  \cite{Ibanez12,dCampo13,Martinez14,Deffner14,Takahashi15} or gauge \cite{Sels17} transformations. Quantum shortcuts to adiabaticity have been extended to non-Hermitian Hamiltonians \cite{Ibanez11,Torosov13}, open quantum systems \cite{Jing13,Vacanti14,Song16,Jing16} and Dirac-dynamics \cite{Deffner16,Muga16,Song17}. They have been demonstrated experimentally \cite{Schaff10,Schaff11,Bason11,Walther12,Zhang13,Kihwan16}, and their relationship with quantum speed limits has been clarified \cite{Campbell17,Funo17}. Shortcuts to adiabaticity are complimentary to, and are sometimes combined with~\cite{Chen11b,Campbell15}, strategies based on optimal quantum control theory, which impose an optimality criterion to improve the fidelity of evolution \cite{Wu15,Glaser15}.

Analogous problems in classical Hamiltonian \cite{Jarzynski13,Deng13,Deffner14,Okuyama16,Patra16,Okuyama17,Jarzynski17} and stochastic \cite{Vaikuntanathan08,Tu14, Martinez16,LeCununder16,Li17} dynamics have recently gained attention. For a classical system in one degree of freedom, evolving under a Hamiltonian $H_0(q,p,t)$, the action $I = \oint p \, dq$ is an adiabatic invariant \cite{Goldstein80}. Classical shortcuts involve constructing an auxiliary Hamiltonian $H_1(q,p,t)$ to  preserve the action even when $H_0(q,p,t)$ changes rapidly. Similarly, the evolution of an overdamped Brownian particle from one thermal equilibrium state to another can be accelerated using an appropriately crafted, time-dependent auxiliary potential.
This approach was developed in Ref.~\cite{Martinez16}, where its similarity with quantum and classical shortcuts was noted.

In this paper, we show that a broad class of quantum, classical and stochastic shortcuts to adiabaticity can be described within a simple, unifying framework. In this framework, the construction of the desired auxiliary Hamiltonian proceeds in three steps. First, we identify the {\it adiabatic evolution} as the evolution that the system of interest would undergo if the process were performed adiabatically. 
We then define velocity and acceleration {\it flow fields} $v(q,t)$ and $a(q,t)$ that characterize the adiabatic evolution, as illustrated in Figs.~\ref{Fig:I_lines}, \ref{Fig:EnergyShell} and \ref{fig:Prob_lines} for the quantum, classical and stochastic cases.
Finally, from these fields we immediately construct auxiliary terms that provide both ``counterdiabatic'' (Eqs.~\ref{eq:qh1cd}, \ref{eq:ch1cd}, \ref{Ucd}) and ``fast-forward'' (Eqs.~\ref{eq:qu1ff}, \ref{eq:cu1ff}) shortcuts.

Our paper is structured as follows. 
In Sec.~\ref{sec:quantum} we develop our approach to quantum shortcuts to adiabaticity, in one degree of freedom.
After deriving the main results of this section, Eqs.~\ref{eq:qh1cd} and \ref{eq:qu1ff}, we compare them with previously obtained shortcuts; we show how they provide insight into the singularities that may arise in the fast-forward approach; we analyze the special case of scale-invariant dynamics; we illustrate our approach with numerical simulations; and we briefly discuss generalizations to three degrees of freedom.
In Sec.~\ref{sec:classical} we review the results of Ref.~\cite{Jarzynski17}, where our approach was applied to classical shortcuts to adiabaticity.
In Sec.~\ref{sec:comparison} we discuss semiclassical connections -- or the lack thereof -- between the quantum and classical results.
In Sec.~\ref{sec:stochastic} we use our approach to construct shortcuts to adiabaticity for an overdamped Brownian particle. 
We present concluding remarks in Sec.~\ref{sec:conclusion}.

%%%%%%%%%%%%%%%%%%%%%%%%%%%%%%%%%%%%%%%%%%%%%%%%%%%%%%%%%%%%%%%%%%%%%%%%%%%%%%%%%%%%%%%%%%%%%%%%%%%%%%%%%%%%%%%%%%%%%%%%%%%%%%%%%%%%%%%%%%%%%%%%%%%%%%%%%%%%%%%%%%%%
\section{Quantum shortcuts}
\label{sec:quantum}

We begin by reviewing two known quantum shortcuts to adiabaticity: {\it transitionless quantum driving} and the {\it fast-forward} method.
The former, due to Demirplak and Rice~\cite{Rice03} and Berry~\cite{Berry09}, involves the counterdiabatic Hamiltonian
\begin{equation}
\label{eq:HCD}
\hat H_{CD}(t) = i\hbar\sum_m \bigl( \vert \partial_t m\rangle \langle m\vert  -  \langle m\vert\partial_t m\rangle \vert m\rangle \langle m \vert \bigr)
\end{equation}
where $\vert m\rangle = \vert m(t)\rangle$ denotes the $m$'th eigenstate of a Hamiltonian of interest, $\hat H_0(t)$, and $\vert \partial_t m\rangle \equiv \partial_t \vert m(t)\rangle$.
If a wavefunction evolves under $\hat H_0 + \hat H_{CD}$ from an initial state $\vert n(0)\rangle$, then it remains in the $n$'th instantaneous eigenstate of $\hat H_0(t)$ at all times, as the term $\hat H_{CD}(t)$ suppresses excitations to other eigenstates~\cite{Rice03,Berry09}.
Note that the counterdiabatic term (Eq.~\ref{eq:HCD}) does not depend on the choice of $n$.
Transitionless quantum driving is derived from basic principles of unitary evolution and is consequently quite general: it applies both to spatially continuous systems such as a particle in a time-dependent potential, and to discrete-state, e.g.\ spin, systems.

The operator $\hat H_{CD}$ (Eq.~\ref{eq:HCD}) is a generator of adiabatic transport~\cite{Jarzynski95,Jarzynski13}, in the sense that
\be
\label{eq:generator}
e^{-i\delta t\,\hat H_{CD}(t)/\hbar} \vert n(t) \rangle = \vert n(t+\delta t) \rangle \, ,
\ee
aside from an overall phase.
Eq.~\ref{eq:generator} clarifies why adding $\hat H_{CD}$ to $\hat H_0$ produces transitionless driving~\cite{Jarzynski13}:
for each time step $\delta t$, the evolution operator under $\hat H_0+\hat H_{CD}$ is $e^{-i\delta t\,\hat H_{0}/\hbar} \, e^{-i\delta t\,\hat H_{CD}/\hbar}$, which first evolves the state $\vert n(t)\rangle$ to $\vert n(t+\delta t)\rangle$, and then contributes an increment in the dynamical phase, $e^{-i\delta t\,E_n/\hbar}$.
Here we have taken $\delta t$ to be infinitesimal, and have ignored ${\cal O}(\delta t^2)$ corrections.

The fast-forward approach, due to Masuda and Nakamura~\cite{Masuda10}, pertains to a Hamiltonian of the form
\begin{equation}
\label{eq:H0_kin+pot}
 \hat H_0(t) = \frac{\hat{\bf p}^2}{2m} + U_0(\hat{\bf q},t) \, .
 \end{equation}
For a given time interval $0\le t\le\tau$, and a particular quantum number $n$, a ``fast-forward'' potential $U_{FF}(\hat{\bf q},t)$ is constructed with the following property: if a wavefunction evolves under $\hat H_0 + \hat U_{FF}$ from the initial state $\vert n(0)\rangle$, then it will arrive in the eigenstate $\vert n(\tau)\rangle$ (up to an overall phase) at $t=\tau$.
For intermediate times $0<t<\tau$, the wavefunction is in a superposition of eigenstates of $\hat H_0(t)$, as illustrated in Fig.~3 of Ref.~\cite{Masuda10}, or Fig.~\ref{Fig:Snapshot} below.
The fast-forward potential depends on the chosen quantum number $n$.
Moreover, $U_{FF}({\bf q},t)$ typically (though not always) becomes singular at nodes of the instantaneous eigenstate, that is, where $\phi_n({\bf q},t) \equiv \langle {\bf q}\vert n(t)\rangle=0$.
Hence the applicability of the fast-forward method is generally restricted to the ground state, $n=0$, although there are exceptions to this statement.
We will return to this point later in our discussion.

Both $\hat H_{CD}$ and $\hat U_{FF}$ are auxiliary Hamiltonians that are tailored to achieve the desired acceleration of adiabatic dynamics.
We will use the term {\it counterdiabatic} ($CD$) to refer to methods in which the auxiliary term causes the system to follow the adiabatic evolution -- at an accelerated pace -- for the duration of the process.
This is the case with transitionless quantum driving: the wavefunction remains in a given eigenstate of $\hat H_0(t)$ at all times, when evolving under $\hat H_0 + \hat H_{CD}$.
The term {\it fast-forward} ($FF$) will refer to methods in which the system strays from the adiabatic evolution at intermediate times, but returns to the adiabatic state at the final time $t=\tau$, as in the Masuda-Nakamura method.
As illustrated by the previous paragraphs, auxiliary terms in the fast-forward approach are {\it local}, in the sense that they are explicit functions of $\hat{\bf q}$ and $t$.
By contrast, counterdiabatic driving generally requires {\it non-local} auxiliary terms, given either by spectral sums (as with Eq.~\ref{eq:HCD}) or by expressions involving both coordinates and momenta (see Refs.~\cite{Jarzynski13,Deffner14,Patra16}, or Eq.~\ref{eq:qh1cd} below).
Thus, fast-forward auxiliary terms may generically be easier to implement experimentally, than counterdiabatic terms.

In what follows, we bridge the two approaches.
We consider a Hamiltonian of interest
\begin{equation}
\label{eq:HofI}
\hat H_0(t) = \frac{\hat p^2}{2m} + U_0(\hat q,t)
\end{equation}
in one degree of freedom.
We assume that $\hat H_0$ varies with time only during the interval $0\le t\le\tau$, and that this time-dependence is turned on and off smoothly --
specifically, $\hat H_0$, $\partial\hat H_0/\partial t$ and $\partial^2\hat H_0/\partial t^2$ are continuous functions of time for all $t$, and $\partial\hat H_0/\partial t=0$ for $t\notin(0,\tau)$.
For a given choice of quantum number $n$, we will define velocity and acceleration flow fields $v(q,t)$ and $a(q,t)$, that characterize how the eigenstate probability distribution $\vert\langle q\vert n(t) \rangle\vert^2$ deforms with $t$.
From these flow fields we will construct simple expressions for both a counterdiabatic Hamiltonian $\hat H_{CD}(\hat q,\hat p,t)$ (Eq.~\ref{eq:qh1cd}), and a local fast-forward potential $\hat U_{FF}(\hat q,t)$ (Eq.~\ref{eq:qu1ff}).

\subsection{Setup and derivation of main results}
\label{subsec:mainResults}

Let the real-valued wavefunction $\phi(q,t) = \langle q \vert n(t) \rangle$ denote the $n$'th eigenstate of $\hat H_0(t)$, with eigenenergy $E(t)$:
\be
\hat H_0(t) \phi(q,t) = 
\left[ -\frac{\hbar^2}{2m} \frac{\partial^2}{\partial q^2} + U_0(q,t) \right] \phi(q,t) = E(t) \phi(q,t).
\label{energyeigen}
\ee
For convenience we omit the subscript $n$ on $\phi$ and $E$.
The {\it adiabatic evolution} is identified as follows:~\footnote{
The dynamical phase $\alpha$ is generically accompanied by a geometric phase~\cite{Berry84}, but the latter vanishes for a kinetic-plus-potential Hamiltonian in one degree of freedom.
}
\be
\psi_{ad}(q,t)=\phi(q,t) \, e^{i \alpha(t)}
\quad,\quad
\alpha(t)= - \frac{1}{\hbar} \int_0^t E(t^\prime)\, dt^\prime
\label{ad_traj}
\ee
When the time-dependence of $\hat H_0(t)$ is quasi-static, $\psi_{ad}$ is a solution of the Schr\" odinger equation, $i\hbar \,\partial_t\psi_{ad} = \hat H_0 \psi_{ad}$~\cite{Born28,Griffiths04}.
When the time-dependence is arbitrary, $\psi_{ad}$ obeys
\be
\label{eq:tdsecd}
i\hbar \frac{\partial\psi_{ad}}{\partial t} = ( \hat H_0+\hat H_{CD} ) \psi_{ad}
\ee
for the counterdiabatic term given by Eq.~\ref{eq:HCD}~\cite{Rice03,Berry09}.
Thus the addition of the term $\hat H_{CD}$ causes the system to follow the adiabatic evolution, $\psi_{ad}$, when $\hat H_0$ is varied rapidly. In what follows we construct a different counterdiabatic term, given as an explicit function of $\hat q$ and $\hat p$ (Eq.~\ref{eq:qh1cd}), which accomplishes the same result for a given choice of $n$.

\begin{figure} 
\centering
\includegraphics[width=0.45\textwidth]{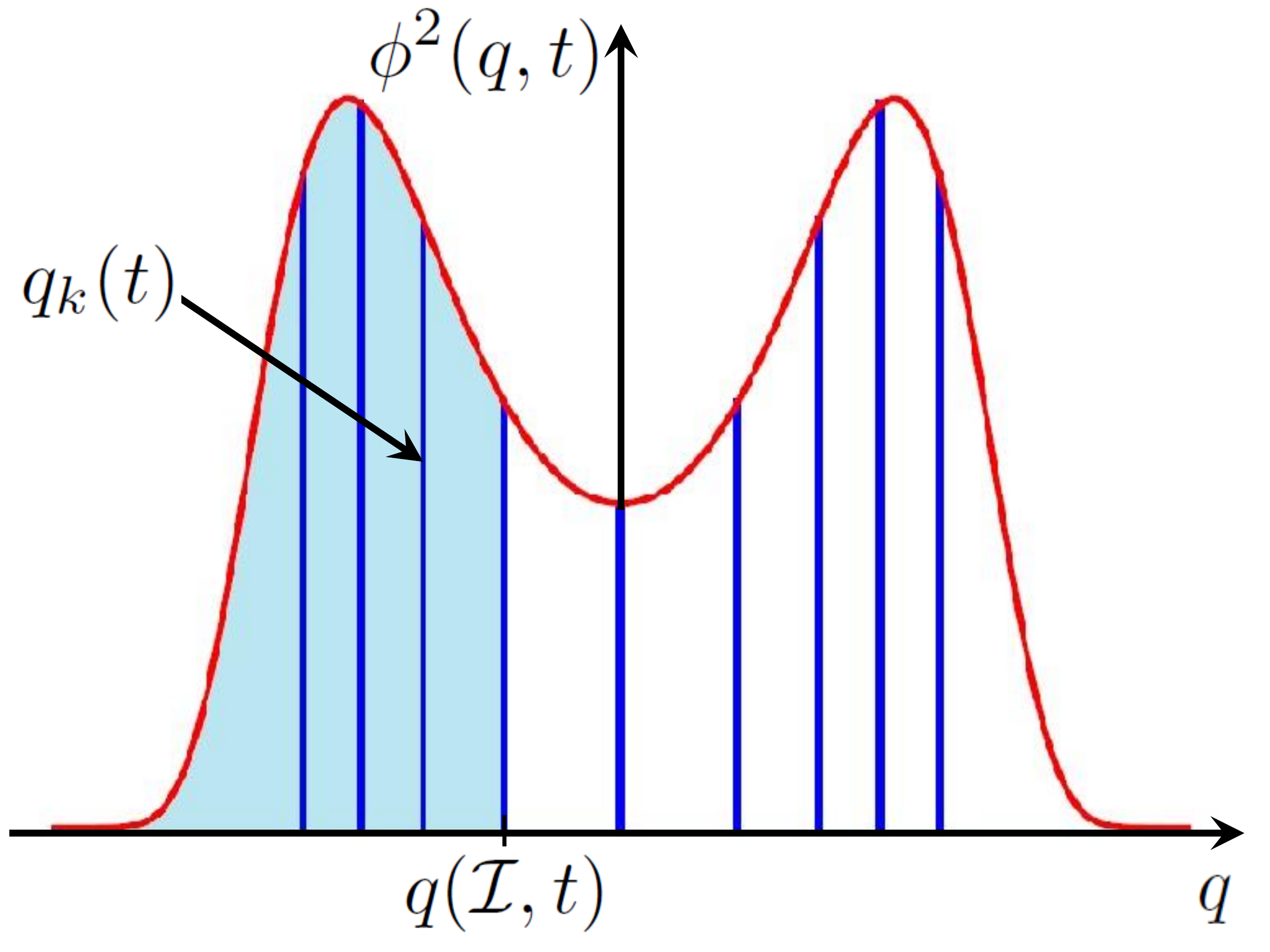}
\caption{\label{Fig:I_lines} %(Color online) 
The red curve $\phi^2(q,t)$ depicts the probability distribution associated with an energy eigenstate of $\hat H_0(t)$. The blue vertical lines divide the area under $\phi^2(q,t)$ into $K\gg 1$ strips of equal area. $q(\mathcal{I},t)$ is the right boundary of the shaded region, which has area $\mathcal{I}$. The positions of the vertical lines vary parametrically with $t$, and this ``motion'' is described in terms of velocity and acceleration fields $v(q,t)$ and $a(q,t)$, as given by Eq.~\ref{flowfqtm}.
}
\end{figure} 

Let us define the integrated probability density function
\be
\mathcal{I}(q,t) \equiv \int_{- \infty}^q  \phi^2(q',t) dq'.
\label{I}
\ee
As $\mathcal{I}(q,t)$ is a monotonically increasing function, we can invert it to obtain $q(\mathcal{I},t)$. We then define a velocity flow field~\footnote{
The quantity $-v(q,t)$ was identified as a ``hydrodynamic velocity'' in Ref.~\cite{Martinez-Garaot16}, Eq.~(6).}
\be
\label{vel}
v(q,t) = \frac{\partial}{\partial t} q(\mathcal{I},t) =  -\frac{\partial_t \mathcal{I}}{\partial_q \mathcal{I}},
\ee
This flow field can be pictured by dividing the area under $\phi^2(q,t)$ into $K\gg 1$ strips of equal area, delimited by vertical lines at locations $q_1(t), q_2(t), \ldots,q_{K-1}(t)$, so that $\mathcal{I}(q_k(t),t) = k/K$; see Fig.~\ref{Fig:I_lines}. The locations $\{ q_k(t) \}$ evolve parametrically with $t$, according to
\begin{subequations}
\label{flowfqtm}
\begin{equation}
\label{eq:vdef}
\frac{dq_k}{dt} = v(q_k,t)
\end{equation}
Note that Eq.~\ref{eq:vdef} does not reflect the unitary dynamics generated by $\hat{H}_0(t)$, but rather the variation of the eigenstate probability density $\phi^2(q,t)$ with $t$.
We similarly introduce an acceleration flow field,
\be
\label{eq:adef}
\frac{d^2q_k}{dt^2} = a(q_k,t)
\ee
\end{subequations}
By Eq.~\ref{eq:vdef} this field satisfies
\be
\label{eq:avvv}
a(q,t) = v^\prime v + \dot v
\ee
where the prime and dot denote $\partial_q$ and $\partial_t$, respectively.
Both flow fields vanish outside the interval $0<t<\tau$:
\begin{equation}
\label{eq:vazero}
v(q,t) = 0 = a(q,t) \quad {\rm for} \,\, t \notin (0,\tau)
\end{equation}
as follows from the assumptions spelled out after Eq.~\ref{eq:HofI}.
We will now use these flow fields to construct counterdiabatic and fast-forward shortcuts, given by Eqs.~\ref{eq:qh1cd} and \ref{eq:qu1ff} below.

We begin by defining the counterdiabatic Hamiltonian,
\be
\label{eq:qh1cd} 
\hat H_{CD}(t) = \frac{\hat p \hat v + \hat v \hat p}{2} \quad,\quad \hat v(t) = v(\hat q,t)
\ee
We claim that the adiabatic wavefunction $\psi_{ad}$ (Eq.~\ref{ad_traj}) satisfies Eq.~\ref{eq:tdsecd}
for arbitrary time-dependence of $\hat H_0(t)$, with $\hat H_{CD}$ now given by Eq.\ref{eq:qh1cd}.
To show this, we first rearrange Eq.~\ref{vel} as $\partial_t \mathcal{I}+v \partial_q \mathcal{I}=0$.
Differentiating both sides with respect to $q$ leads to the continuity equation
$\partial_t \phi^2 + \partial_q(v \phi^2)=0$,
equivalently,
\be
\dot\phi + v \phi^\prime+\frac{1}{2} v^\prime \phi =0.
\label{cont2} 
\ee
We now use Eqs.~\ref{energyeigen}, \ref{ad_traj} and \ref{cont2} to evaluate the right side of Eq.~\ref{eq:tdsecd}:
\ba
( \hat H_0+ \hat H_{CD} ) \psi_{ad}
&=&\left[ \hat H_0 \phi -\frac{i \hbar}{2}\partial_q(v \phi)- \frac{i \hbar}{2}v (\partial_q \phi)  \right] e^{i \alpha} \non \\
&=& \left[E \phi -i \hbar\left(v \phi^\prime+\frac{1}{2} v^\prime \phi \right) \right] e^{i \alpha} \non \\
&=& \left( E \phi + i \hbar\, \dot\phi \right)e^{i \alpha} = i \hbar\, \frac{\partial\psi_{ad}}{\partial t}
\ea
which establishes that $\psi_{ad}(q,t)$ is a solution of Eq.~\ref{eq:tdsecd}.
Thus if a wavefunction evolves under $\hat H_0 + \hat H_{CD}$ from an initial state $\psi(q,0) =  \langle q \vert n(0) \rangle$, then it remains in the $n$'th instantaneous eigenstate of $\hat H_0(t)$ during the entire process, just as in the case of transitionless quantum driving~\cite{Rice03,Berry09}.

Turning our attention to the fast-forward approach, we construct a potential $U_{FF}(q,t)$ and a companion function $S(q,t)$ as follows:
\ba
\label{eq:qu1ff}
-\partial_q U_{FF} &=& ma(q,t) \\
\label{eq:S}
\partial_q S &=& mv(q,t)
\ea
By Eq.~\ref{eq:avvv}, these functions satisfy
\ba
\label{eq:preHJ}
\partial_q \left[ \partial_t S + \frac{1}{2m} (\partial_q S)^2  + U_{FF}\right] = 0
\ea
Eqs.~\ref{eq:qu1ff} and \ref{eq:S} specify $U_{FF}$ and $S$ only up to arbitrary functions of time.
We use this freedom, along with Eqs.~\ref{eq:vazero} and \ref{eq:preHJ}, to impose the conditions
\be
\label{eq:UFF_on_off}
U_{FF}(q,t) = 0 \quad \text{for} \,\, t \notin (0,\tau)
\ee
and
\be
\label{eq:UFF_HJ}
\partial_t S+\frac{1}{2m} (\partial_q S)^2 + U_{FF} = 0
\ee
Eq.~\ref{eq:vazero} further implies that
\be
\label{eq:Spm}
S(q,0) = S_- \quad,\quad
S(q,\tau) = S_+
\ee
where $S_\pm$ are constants (i.e.\ independent of $q$).

We now show that the ansatz
\be
\label{eq:ansatz}
\bar{\psi}(q,t)=\psi_{ad}(q,t) \exp \left[i \frac{S(q,t)}{\hbar}\right] = \phi\, e^{i\alpha} e^{iS/\hbar}
\ee
is a solution of the Schr\" odinger equation
\be
i\hbar \frac{\partial\bar\psi}{\partial t} = ( \hat H_0+\hat U_{FF} ) \bar\psi
\ee
Evaluating the right side with the help of Eqs.~\ref{energyeigen}, \ref{cont2}, \ref{eq:S} and \ref{eq:UFF_HJ}, we obtain
\ba
&& \left( -\frac{\hbar^2}{2m} \frac{\partial^2}{\partial q^2} + U_0 + U_{FF} \right) \phi(q,t)\, e^{i\alpha}\, e^{iS/\hbar} \non \\
&=&
\left[
-\frac{\hbar^2}{2m} \phi^{\prime\prime} - \frac{i\hbar}{m} \phi^\prime S^\prime - \frac{i\hbar}{2m} \phi S^{\prime\prime} + \frac{1}{2m} \phi (S^\prime)^2 + U_0\phi + U_{FF}\phi
\right] \, e^{i\alpha}\, e^{iS/\hbar} \non \\
&=&
\left[
E\phi + \frac{1}{2m} (S^\prime)^2 \phi + U_{FF} \phi - i\hbar \left( v \phi^\prime + \frac{1}{2} v^\prime \phi \right)
\right] \, e^{i\alpha}\, e^{iS/\hbar} \non \\
&=&
\left( i\hbar\dot\phi + E\phi - \dot S\phi \right) \, e^{i\alpha}\, e^{iS/\hbar}
= i\hbar \frac{\partial\bar\psi}{\partial t}
\ea
which is the desired result.
By Eq.~\ref{eq:Spm}, the wave function $\bar\psi(q,t)$ begins in the $n$'th energy eigenstate at $t \le 0$ and ends in the $n$'th energy eigenstate at $t \ge\tau$, which establishes that $\hat U_{FF}$ produces fast-forward evolution.

Note that we introduced the function $S(q,t)$ (Eq.~\ref{eq:S}) only to facilitate the derivation of our fast-forward approach.
This function need not be evaluated if one simply wishes to construct the potential $U_{FF}(q,t)$.
That potential can be determined directly from the acceleration field $a(q,t)$, by Eq.~\ref{eq:qu1ff}.

Also, we imposed Eq.~\ref{eq:UFF_on_off} so as to obtain an auxiliary potential that is turned on at $t=0$ and off at $t=\tau$, but this condition is not necessary.
Any $U_{FF}$ satisfying Eq.~\ref{eq:qu1ff} will provide a shortcut that transports the $n$'th eigenstate of $\hat H_0(0)$ to the $n$'th eigenstate of $\hat H_0(\tau)$.
The addition of an arbitrary function $f(t)$ to $U_{FF}(q,t)$ affects only the overall phase of the evolving wavefunction.

\subsection{Comparison with previous results}
\label{subsec:qcompare}

Eqs.~\ref{eq:qh1cd} and \ref{eq:qu1ff} are recipes for constructing shortcuts directly from the flow fields $v(q,t)$ and $a(q,t)$.
Let us compare these results with previously published counterdiabatic and fast-forward shortcuts.

Our result for $\hat H_{CD}$ (Eq.~\ref{eq:qh1cd}) is given explicitly in terms of the operators $\hat q$ and $\hat p$.
This appealing feature comes with a cost: in  general, a different counterdiabatic term is required for each eigenstate $n$, since $v(q,t)$ depends on the choice of $n$.
By contrast the Demirplak-Rice-Berry~\cite{Rice03,Berry09} counterdiabatic term (Eq.~\ref{eq:HCD}) is independent of $n$, as noted earlier.
We conclude that Eqs.~\ref{eq:HCD} and \ref{eq:qh1cd} are not equivalent, although the two counterdiabatic terms produce the same effect when they act on the chosen adiabatic eigenstate:
\be
\label{eq:but}
\hat H_{CD}^{{\rm Eq.}\ref{eq:HCD}} \ne \hat H_{CD}^{{\rm Eq}.\ref{eq:qh1cd}}
\quad \textrm{but} \quad
\hat H_{CD}^{{\rm Eq.}\ref{eq:HCD}} \vert n\rangle = \hat H_{CD}^{{\rm Eq}.\ref{eq:qh1cd}} \vert n\rangle
\ee
Using the identity $\langle m\vert\partial_t m\rangle = 0$, which holds for $\hat H_0$ given by Eq.~\ref{eq:HofI}, we rewrite the equality in Eq.~\ref{eq:but} as follows:
\be
i\hbar \, \frac{\partial}{\partial t} \vert n \rangle = \frac{\hat p \hat v + \hat v \hat p}{2} \, \vert n \rangle
\ee
In other words, the operator $\hat D \equiv (\hat p \hat v + \hat v \hat p)/2$ acts as the generator of adiabatic transport (see Eq.~\ref{eq:generator}) for the state $\vert n\rangle$ that was used to construct $v(q,t)$:
\be
e^{-i\delta t\hat D/\hbar} \vert n(t) \rangle = \vert n(t+\delta t) \rangle
\ee
for infinitesimal $\delta t$. 

Substituting Eq.~\ref{eq:S} into Eq.~\ref{cont2} yields
\be
\dot\phi+\frac{1}{m} S^\prime \phi^\prime +\frac{1}{2m}\phi S^{\prime\prime}=0.
\label{eq:S_simp}
\ee
Eqs.~\ref{eq:UFF_HJ} and \ref{eq:S_simp} are essentially equivalent to Eqs.~5 and 6 of Torrontegui {\it et al}~\cite{Torrontegui12}, to Eqs.~17 and 15 of Takahashi~\cite{Takahashi15}, and to Eqs.~4 and 3 of Mart\' inez-Garaot {\it et al}~\cite{Martinez-Garaot16}.
In Refs.~\cite{Torrontegui12,Takahashi15,Martinez-Garaot16} these equations were used to provide streamlined derivations of the fast-forward approach pioneered by Masuda and Nakamura~\cite{Masuda10}.
(Our Eq.~\ref{eq:S_simp} is also equivalent to Eq.~2.18 of Ref.~\cite{Masuda10}, and our Eq.~\ref{eq:S} appears as Eq.~5 in Ref.~\cite{Martinez-Garaot16}.)
Thus our fast-forward potential $\hat U_{FF}$ is equivalent to that derived by previous authors~\cite{Masuda10,Torrontegui12,Takahashi15,Martinez-Garaot16}.

The observation that the quantum counterdiabatic and fast forward approaches are closely related is not surprising, as previous authors have argued that $\hat U_{FF}$ can be constructed from $\hat H_{CD}$ by appropriate unitary  \cite{Ibanez12,dCampo13,Martinez14,Deffner14,Takahashi15} or gauge \cite{Sels17} transformations.
The novelty of our approach is that we obtain both $\hat H_{CD}$ and $\hat U_{FF}$ directly from the velocity and acceleration fields that describe the time-dependence of $\phi^2(q,t)$ (Fig.~\ref{Fig:I_lines}).
Our results are given by compact, intuitive expressions (Eqs.~\ref{eq:qh1cd}, \ref{eq:qu1ff}).
This approach highlights the connection between counterdiabatic and fast-forward shortcuts, and -- as we shall see -- it provides insight into the divergences that often plague the fast-forward method when it is applied to excited states.
Moreover, the construction of $\hat H_{CD}$ and $\hat U_{FF}$ from $v$ and $a$ is mirrored in classical shortcuts to adiabaticity, as will be discussed in Sec.~\ref{sec:classical}.

Finally, we note that Eq.~\ref{eq:UFF_HJ} is a Hamilton-Jacobi equation for the Hamiltonian  $p^2/2m+U_{FF}$.
Okuyama and Takahashi~\cite{Okuyama17} have recently used the Hamilton-Jacobi formalism to explore the correspondence between quantum and classical shortcuts to adiabaticity.
It would be interesting to explore the relationship between their approach and ours.

\subsection{Divergences and a ``no-flux'' criterion}
\label{subsec:divergences}

By Eq.~\ref{vel}, $v(q,t)$ generically diverges at nodes of the wavefunction, where $\partial_q \mathcal{I} = \phi^2$ vanishes; this in turn leads to divergences in $a(q,t)$, and in $\hat H_{CD}$ and $\hat U_{FF}$.
These observations suggest that our method is in general restricted to ground state wavefunctions ($n=0$), which have no nodes.

While nodes in $\phi(q,t)$ {\it typically} spoil the applicability of our method, this need not always be the case: the numerator and denominator in Eq.~\ref{vel} might vanish simultaneously in a way that prevents the ratio $v = -\partial_t\mathcal{I}/\partial_q\mathcal{I}$ from blowing up at a node.
Here we propose a simple criterion for determining whether our approach (and by extension the fast-forward approach~\cite{Masuda10,Torrontegui12,Takahashi15,Martinez-Garaot16}) is applicable when an eigenstate $\phi(q,t)$ has one or more nodes.

Let $q_\nu(t)$ denote the location, and $u_\nu(t) \equiv dq_\nu/dt$ the velocity, of the $\nu$'th node of $\phi(q,t)$.
We assume $\vert u_\nu\vert < \infty$, as will generally be the case when the potential $U_0(q,t)$ is well-behaved.
As $t$ varies parametrically, the flux of probability across this node, from the region $q<q_\nu$ to the region $q>q_\nu$, is given by
\be
\label{eq:flux}
\Phi_\nu(t) = -\frac{d}{dt} \mathcal{I}(q_\nu,t) = \left[ v(q_\nu,t) - u_\nu(t) \right] \phi^2(q_\nu,t)
\ee
using Eqs.~\ref{I} and \ref{vel}.
This result has the familiar interpretation of ``flux equals velocity times density'', in the node's co-moving frame of reference.
Eq.~\ref{eq:flux} implies that if $v(q,t)$ does not blow up at a given node, then the probability flux $\Phi_\nu(t)$ across that node must be zero.
This suggests a simple criterion: if the time-dependence of $\phi^2(q,t)$ is such that there is no flux of probability across any node, i.e.\ if $\Phi_\nu=0$ for all $\nu$, then the velocity field $v(q,t)$ will not diverge at the nodes and our method will remain valid and applicable~\footnote{
In fact, if $\Phi_\nu(t)=0$ then $v(q_\nu,t)=u_\nu(t)$, although we will not make use of this result here.
}.
Generalizing the term ``nodes'' to include the boundaries at $q = \pm\infty$, the no-flux criterion can alternatively be stated as follows:
if the probability between all pairs of adjacent nodes remains independent of $t$
[i.e.\ if $(d/dt)\int_{q_\nu}^{q_{\nu+1}} \phi^2 dq=0$ for all $\nu$],
then $v(q,t)$ will be free of divergences and $U_{FF}(q,t)$ will be well-behaved.

This ``no-flux'' criterion is {\it not} generically satisfied for $\hat H_0(t)$ given by Eq.~\ref{eq:HofI}.
However, in Sec.~\ref{subsec:scaleInvariant} we consider a particular class of time-dependent Hamiltonians for which this criterion is satisfied for every eigenstate, due to {\it scale-invariance} (Eq.~\ref{SI_H0}).
In agreement with the arguments presented above, our method provides non-singular counterdiabatic and fast-forward shortcuts for all energy eigenstates, for this class of Hamiltonians.
In Sec.~\ref{subsec:model} we present the results of numerical simulations for a {\it non}-scale-invariant Hamiltonian, for which the no-flux criterion is satisfied for the first excited state; again, our method successfully provides a non-singular shortcut for this situation.

Divergences associated with eigenstate nodes are problematic not only for our approach, but also for those of Refs.~\cite{Masuda10,Torrontegui12,Takahashi15,Martinez-Garaot16}, since all these approaches lead to equivalent expressions for $U_{FF}$.
This problem has not received much attention in the literature, although Mart\' inez-Garaot {\it et al}~\cite{Martinez-Garaot16} consider it in a slightly different context.
In Sec.\ III.D of their paper, they develop a fast-forward strategy to drive a wavefunction from a ground state $\phi_0$ to a first excited state $\phi_1$.
In their approach the fast-forward potential becomes singular due to the node in $\phi_1$, but they demonstrate that {\it ad hoc} truncation of the singularity produces a well-behaved potential that achieves near-perfect fidelity.
It would be interesting to test whether such truncation is also useful in the context of our method, when the no-flux criterion is not satisfied.

\subsection{Scale-invariant dynamics}
\label{subsec:scaleInvariant}

In the special case of {\it scale-invariant driving}, $U_0(q,t)$ undergoes expansions, contractions and translations.
As shown in Ref.~\cite{Deffner14} (and anticipated in Refs.~\cite{Lewis82,Muga10,dCampo12a,DiMartino13,dCampo13,Jarzynski13,Deng13}), simple expressions for counterdiabatic and fast-forward shortcuts can be obtained when a system is driven in a scale-invariant manner.
In this section we show that these shortcuts are obtained naturally within our framework.

The Hamiltonian for scale-invariant driving takes the form \cite{Deffner14}
\be
\hat{H}_0(t) = \hat H_0(\gamma,f) = \frac{\hat{p}^2}{2m} +\frac{1}{\gamma ^2}U_0 \left(\frac{\hat q-f}{\gamma}\right),
\label{SI_H0}
\ee
where $\gamma(t)$ and $f(t)$ are parameters that describe expansions/contractions, and translations, respectively.
If we let $\tilde{\phi}(q)$ denote the $n$'th eigenstate of $\hat{H}_0(\gamma=1,f=0)$, then the $n$'th eigenstate for a general choice of $(\gamma,f)$ is given by~\cite{Deffner14}
\be
\label{eq:scaling}
\phi(q)= \frac{1}{\sqrt{\gamma}} \tilde{\phi}\left(\frac{q-f}{\gamma}\right)
\ee
This scaling result immediately reveals how the ``picket fence'' of lines $\{ q_k \}$ depicted in Fig.~\ref{Fig:I_lines} behaves when $\gamma$ and $f$ are varied with time.
Variations in $f$ result in translations of the entire picket fence, and variations in $\gamma$ cause the picket fence to expand or contract linearly.
These considerations give us
\be
v(q,t)=\frac{\dot{\gamma}}{\gamma}(q-f) + \dot{f}, 
\label{SI_vel}
\ee
and therefore (by Eq.~\ref{eq:avvv})
\be
a(q,t) = \frac{\ddot{\gamma}}{\gamma} (q-f) + \ddot{f} .
\label{SI_accn}
\ee
Eq.~\ref{SI_vel} also follows from Eq.~\ref{vel}, $v = -\partial_t\mathcal{I}/\partial_q\mathcal{I}$, since $\partial_q \mathcal{I} = \phi^2$ and (making use of Eq.~\ref{eq:scaling})
\ba
\partial_t \mathcal{I}
&=& \left( \dot{\gamma} \, \partial_{\gamma} + \dot{f} \, \partial_{f} \right) \int_{-\infty}^q \frac{1}{\gamma}\tilde{\phi}^2\left(\frac{q'-f}{\gamma}\right) \, dq' \non \\
&=& \dot{\gamma} \int_{-\infty}^q  - \frac{1}{\gamma^2} \left[ \tilde{\phi}^2 + 2(q'-f) \tilde{\phi} (\partial_{q'}\tilde{\phi}) \right]  dq'
+ \dot f \int_{-\infty}^q -\frac{2\tilde{\phi} (\partial_{q'}\tilde{\phi})}{\gamma} dq'
\non\\
&=& 
-\frac{\dot{\gamma}}{\gamma} \int_{-\infty}^q \left[ \phi^2 + (q'-f) \partial_{q'}(\phi^2) \right] dq' - \dot{f} \int_{-\infty}^q \partial_{q'}(\phi^2) dq'
\non\\
&=& \left[ -\frac{\dot{\gamma}}{\gamma} (q-f) -\dot{f} \right] \phi^2
\ea

Combining Eqs.~\ref{SI_vel} and \ref{SI_accn} with Eqs.~\ref{eq:qh1cd} and \ref{eq:qu1ff}, we obtain
\begin{subequations}
\label{eq:SI_shortcuts}
\ba
\label{SI_H1cd}
\hat H_{CD} &=& \frac{\dot{\gamma}}{2 \gamma} \left[ (\hat{q}-f) \hat{p} + \hat{p} (\hat{q}-f) \right] +\dot{f}\hat{p} \\
\label{SI_Hiff}
\hat U_{FF} &=& -\frac{m}{2}\frac{\ddot{\gamma}}{\gamma} (\hat q - f)^2  - m \ddot{f} \hat q
\ea
\end{subequations}
in agreement with Eqs.\ 9 and 30 of Ref.~\cite{Deffner14}.
Thus the shortcuts obtained previously for scale-invariant driving emerge naturally within our framework, from the flow fields $v$ and $a$ (Eqs.~\ref{SI_vel}, \ref{SI_accn}).

We end this section by highlighting two exceptional features of scale-invariant Hamiltonians, both of which are due to the fact that all of the eigenstates of $\hat H_0$ satisfy the same scaling property, Eq.~\ref{eq:scaling}.
First, although $\phi(q) = \langle q\vert n\rangle$ denotes a specific energy eigenstate in the above calculations, the resulting flow fields and shortcuts (Eq.~\ref{eq:SI_shortcuts}) are independent of the choice of $n$.
This suggests that $\hat H_{CD}^{{\rm Eq.}\ref{eq:HCD}} = \hat H_{CD}^{{\rm Eq}.\ref{eq:qh1cd}}$ for scale-invariant driving, in contrast with the general situation discussed in Sec.~\ref{subsec:qcompare}.
Indeed, it has been shown elsewhere that Eq.~\ref{SI_H1cd} -- which we derived from Eq.~\ref{eq:qh1cd} -- follows directly from Eq.~\ref{eq:HCD}~\cite{Deffner14}.
Secondly, the shortcuts given by Eq.~\ref{eq:SI_shortcuts} do not suffer from divergences at the nodes of excited energy eigenstates.
This is easy to understand in terms of the no-flux criterion of Sec.~\ref{subsec:divergences}:
because variations in $\gamma$ merely cause the eigenstate $\phi$ to expand or contract linearly, and variations in $f$ induce simple translations of $\phi$, the probability between adjacent nodes of the wavefunction is independent of $t$.

\subsection{Numerical illustration of fast-forward driving}
\label{subsec:model}

\begin{figure} 
\centering
\includegraphics[width=0.45\textwidth]{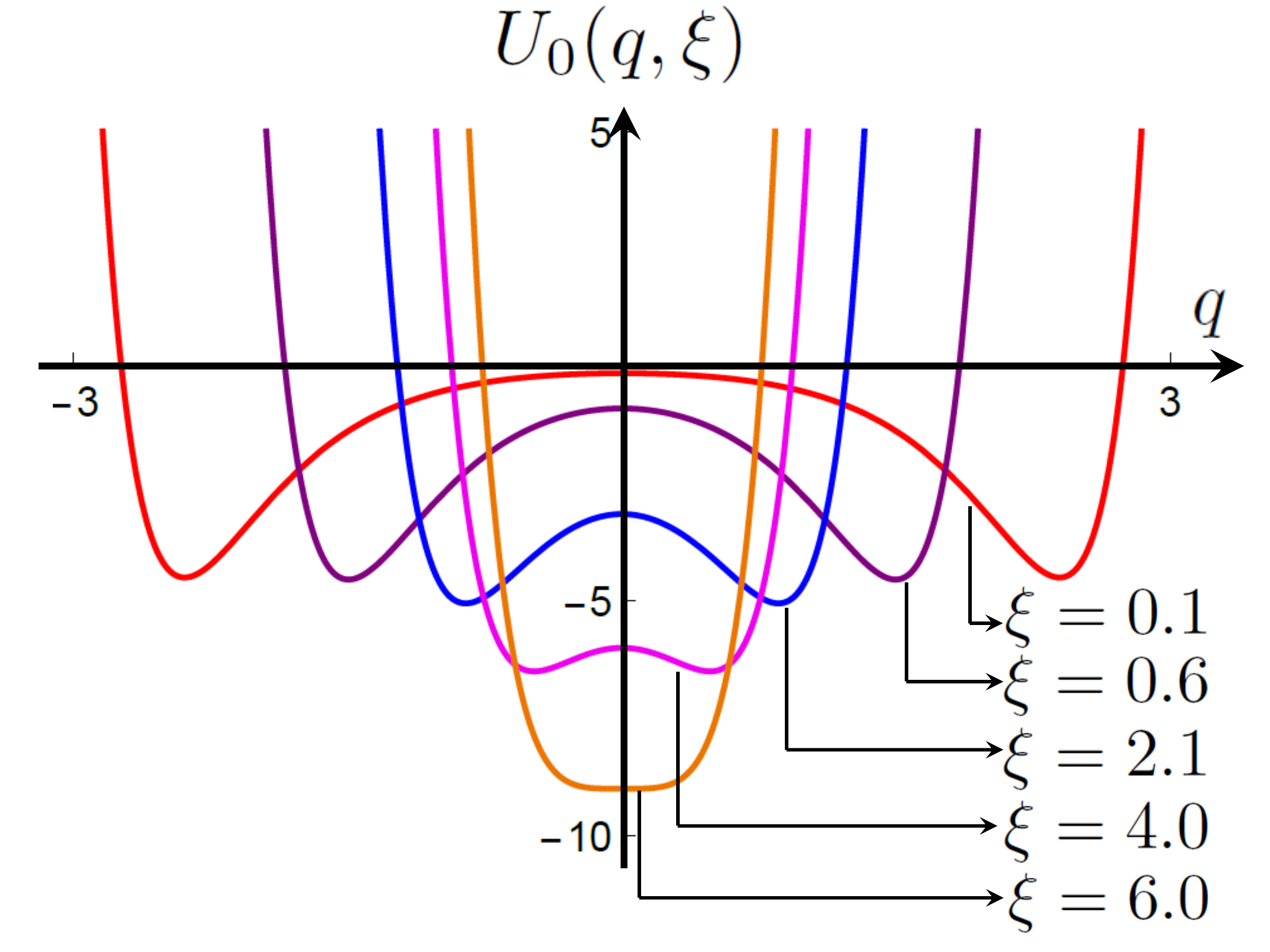}
\caption{\label{Fig:Pot1} %(Color online) 
$U_0(q,\xi)$ is plotted for five values of $\xi$.}
\end{figure}

The parameter-dependent potential
\be
U_0(q;\xi) = \frac{\xi ^2}{16}  \left[ \cosh (4 q)-1 \right] - \frac{3\xi}{2} \cosh (2 q) ,
\label{PotRazavy}
\ee
belongs to a class of potentials studied by Razavy~\cite{Razavy80}, for which convenient analytical expressions for low-lying eigenstates can be obtained.
Here and below, we have set the quantities $\beta$, $m$ and $\hbar$ (appearing in Ref.~\cite{Razavy80}) to unity, so as to work with an effectively dimensionless Hamiltonian.
As illustrated in Fig.~\ref{Fig:Pot1}, $U_0(q;\xi)$ changes from a broad double well to a narrow single well as $\xi$ is increased from 0.1 to 6.0.

Now consider
\ba 
\hat{H}_0(t) = \frac{\hat{p}^2}{2}+ U_0(\hat{q};\xi(t))
\label{Hamil_DW}
\ea
where $\xi(t)$ varies monotonically from $0.5$ to $8.5$ over the interval $0 \le t \le \tau$, according to
\be
\xi(t) = 4.5 + \cos\left(\frac{\pi t}{\tau}\right) \left[ \cos\left(\frac{2\pi t}{\tau}\right) - 5\right] ,
\label{xivar}
\ee
and $\xi(t)$ remains constant outside this interval.
Note that $\dot\xi(0)=\dot\xi(\tau)=0$ and $\ddot\xi(0)=\ddot\xi(\tau)=0$, hence $\hat H_0(t)$ satisfies the continuity conditions described after Eq.~\ref{eq:HofI}.

The wavefunction for the first excited state of $\hat H_0(t)$ is given by \cite{Razavy80}
\be
\phi(q,t) = \kappa(t) \sinh(2q) \exp \left[ - \frac{1}{4} \xi(t) \cosh (2 q) \right] 
\label{psi_razavy}
\ee
where $\kappa(t)$ is set by normalization.
The corresponding eigenenergy is $E(t)=-2$.
Although this eigenstate has a node at the origin, the no-flux criterion of Sec.~\ref{subsec:divergences} is satisfied by the anti-symmetry of the wavefunction:
$\phi(-q,t) = -\phi(q,t)$, hence $\mathcal{I}(0,t) = 1/2$ for all $t$.
Thus we expect our approach to apply despite the presence of the node.

Using the above expressions and setting $\tau = 0.2$, we numerically computed the function $\mathcal{I}(q,t)$ (Eq.~\ref{I}), from which we constructed the flow fields $v$ and $a$ and the fast-forward potential $U_{FF}$ (Eq.~\ref{eq:qu1ff}).
As $\xi(t)$ increases from $0.5$ to $8.5$, $U_0(q,t)$ becomes increasingly narrow (Fig.~\ref{Fig:Pot1}), as does the eigenstate $\phi(t)$; this is reflected in the fields $v$ and $a$, which describe the flow of probability toward the origin.
$U_{FF}(q,t)$ initially develops into a potential well that resembles a parabola (though it is not precisely quadratic) -- this brings about the acceleration of probability flow toward the origin.
%In particular, while $U_{FF}=0$ at $t=0$, it soon forms a potential well that resembles a parabola (though it is not perfectly quadratic) -- this brings about the acceleration of probability flow toward the origin.
At later times during the process $U_{FF}$ resembles an {\it inverted} parabola, which effectively decelerates this flow.

We performed two numerical simulations of evolution under the time-dependent Schr\"odinger equation
\be
i\hbar\frac{\partial\psi}{\partial t} = \hat H(t) \psi
\ee
using $\hat H = \hat H_0$ in the first simulation and $\hat H = \hat H_0 + \hat U_{FF}$ in the second; we will use the notation $\psi^0$ and $\psi^{FF}$ to distinguish between the two simulations.
In both cases the wavefunction was initialized in the state $\psi(q,0) = \phi(q,0)$.
The time evolution was performed using the split-time propagation scheme~\cite{Feit82,Kosloff88}, which involves the repeated application of the fast Fourier transform to toggle between the position and momentum representations.

\begin{figure*}
\centering
\includegraphics[width=0.95\textwidth]{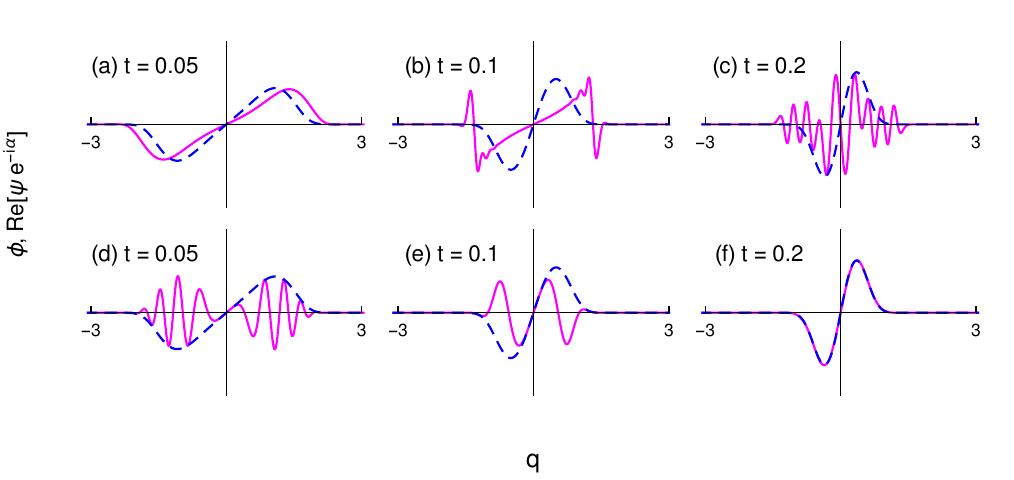}
\caption{\label{Fig:Snapshot_ES} %(Color online)
Evolution under $\hat H_0$ (upper panel) and $\hat H_0+\hat U_{FF}$ (lower panel).
The solid magenta lines show ${\rm Re}(\psi e^{-i\alpha})$, and the dashed blue lines show the eigenstate $\phi$.
Snapshots are shown at $t=0.05$, at $t=0.1$, and at the end of the process, $t=0.2$.
}
 \label{Fig:Snapshot}
\end{figure*} 

Fig.~\ref{Fig:Snapshot_ES} shows snapshots of $\psi(q,t)e^{-i\alpha(t)}$ (solid lines) and $\phi(q,t)$ (dashed lines) for both simulations.
Note that $\psi e^{-i\alpha} = \phi$ in the adiabatic limit (Eq.~\ref{ad_traj}) --
this is our motivation for plotting $\psi e^{-i\alpha}$ rather than $\psi$, though in the following paragraphs we largely will stop writing the factor $e^{-i\alpha}$, for convenience.

The upper panel of Fig.~\ref{Fig:Snapshot_ES} shows the evolution of $\psi^0(q,t)$.
Due to the nonadiabatic time-dependence of $\hat H_0$, the wavefunction $\psi^0$ ``lags'' behind the instantaneous eigenstate $\phi$.
This is particularly evident in Fig.~\ref{Fig:Snapshot_ES}(b), where the probability associated with $\phi$ has shifted substantially toward the origin, while $\psi^0$ remains somewhat behind.
This lag leads to shock waves, which are nascent in Fig.~\ref{Fig:Snapshot_ES}(b).
These shocks propagate inward, and $\psi^0$ ends in a superposition of excited states [Fig.~\ref{Fig:Snapshot_ES}(c)].

The lower panel shows the evolution of $\psi^{FF}(q,t)$.
Here the wavefunction develops excitations at short times [Fig.~\ref{Fig:Snapshot_ES}(d)], in response to large forces generated by $\hat U_{FF}(t)$.
These forces eliminate the lag that is observed in the upper panel, by ``squeezing'' the wavefunction and causing probability to accelerate toward the origin.
At later times this flow is decelerated -- again, due to $\hat U_{FF}(t)$ -- and the excitations subside [Fig.~\ref{Fig:Snapshot_ES}(e)].
The wavefunction gently arrives at the desired energy eigenstate at the final time [Fig.~\ref{Fig:Snapshot_ES}(f)].

In the present context Eq.~\ref{eq:ansatz} can be written as
\be
\label{eq:psiFF}
\psi^{FF} e^{-i\alpha} = \phi\, e^{iS/\hbar}
\ee
which implies that the probability densities $\vert\psi^{FF}\vert^2 = \vert\phi\vert^2$ at all times, despite the excitations that develop in $\psi^{FF}(q,t)$.
We have verified this result in our simulations (data not shown).
Eq.~\ref{eq:psiFF} further implies that ${\rm Re}(\psi^{FF}e^{-i\alpha}) = \phi(q,t)\cos[S(q,t)/\hbar]$, which is illustrated in Fig.~\ref{Fig:Snapshot_ES}(d), where the dashed line is manifestly the envelope of the solid line.

\begin{figure}
\centering
\includegraphics[width=0.35\textwidth]{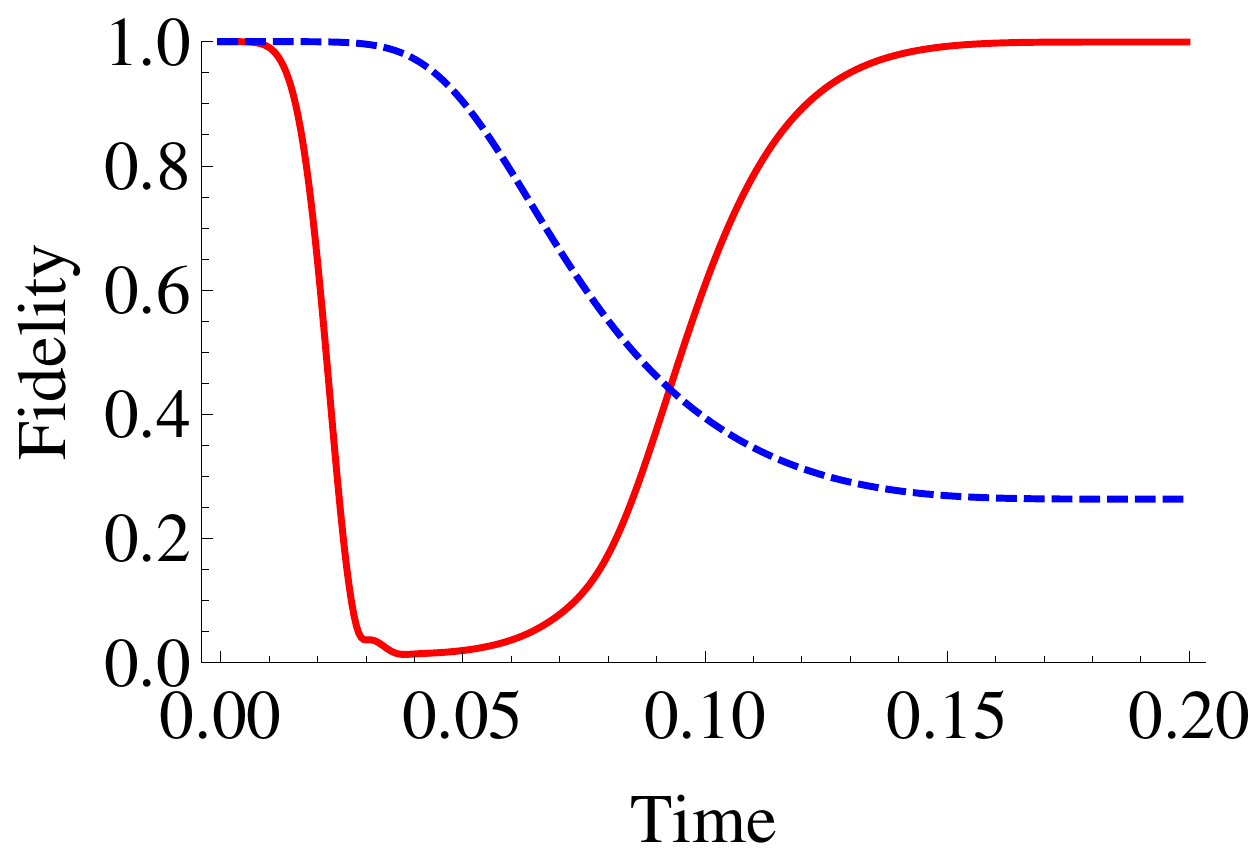}
\caption{ %(Color online) 
The blue dashed curve shows the fidelity $\vert\langle\phi\vert\psi^0\rangle\vert^2$, quantifying the limited extent to which $\psi^0(q,t)$, evolving under $\hat H_0$, keeps pace with the energy eigenstate $\phi(q,t)$.
The solid red curve shows $\vert\langle\phi\vert\psi^{FF}\rangle\vert^2$, which is the fidelity that is achieved when $\hat U_{FF}$ is added to the Hamiltonian.}
\label{Fig:Fid21}
\end{figure} 

Finally, for both simulations we computed the fidelity $F(t) = \vert\langle\phi\vert\psi\rangle\vert^2$, that is the degree of overlap between the evolving wavefunction and the energy eigenstate.
Fig.~\ref{Fig:Fid21} shows the results.
In the absence of the auxiliary term, the fidelity $\vert\langle\phi\vert\psi^0\rangle\vert^2$ decays monotonically to $F \approx 0.3$.
When $\hat U_{FF}$ is included in the Hamiltonian, the fidelity $\vert\langle\phi\vert\psi^{FF}\rangle\vert^2$ at first drops rapidly to nearly zero -- due to the excitations that develop in $\psi^{FF}$ -- but then it claws its way back to unity, illustrating the effectiveness of the fast-forward potential obtained from the acceleration flow field $a(q,t)$.

\subsection{Extension to three degrees of freedom}
\label{subsec:3d}

Although the focus in this paper is on systems with one degree of freedom, here we briefly discuss how the results of Sec.~\ref{subsec:mainResults} might be extended to three dimensions.
We will use boldface to denote vector quantities.

For a given choice of quantum number $n$, let $\phi({\bf q},t)$ and $E(t)$ denote the $n$'th eigenstate and eigenenergy, respectively, of the Hamiltonian $\hat H_0(t)$ given by Eq.~\ref{eq:H0_kin+pot}.
Let us define a vector field ${\bf v}({\bf q},t)$ by the equation
\be
\label{eq:continuity3d}
\partial_t \phi^2 + \nabla\cdot({\bf v}\phi^2)=0
\ee
which describes how the eigenstate probability density $\phi^2({\bf q},t)$ varies parametrically with $t$.
We assume that $\hat H_0(t)$ and its first two time derivatives vanish outside the interval $0<t<\tau$ (as in the one-dimensional case), therefore ${\bf v}$ can be constructed to vanish outside this interval as well: ${\bf v} = {\bf 0}$ for $t\notin(0,\tau)$.~\footnote{
Eq.~\ref{eq:continuity3d} defines ${\bf v}({\bf q},t)$ only up to gauge-like transformations of the form ${\bf v} \rightarrow {\bf v} + (\nabla\times{\bf B})/\phi^2$, where ${\bf B}({\bf q},t)$ is an arbitrary, well-behaved vector field.
Hence we have some freedom in constructing ${\bf v}$.
This freedom was not present in Sec.~\ref{subsec:mainResults}, where the $v(q,t)$ was defined using the construction shown in Fig.~\ref{Fig:I_lines}, rather than from the continuity equation.
}

Since Eq.~\ref{eq:continuity3d} is a continuity equation, it can be interpreted as describing an ensemble of independent trajectories, each evolving according to $\dot{\bf q} = {\bf v}({\bf q},t)$.
The acceleration of these trajectories is described by a field ${\bf a}({\bf q},t)$ whose $i$'th component satisfies
\be
\label{eq:ai3d}
a_i = \ddot q_i = \frac{\partial v_i}{\partial t} + \sum_j \frac{\partial v_i}{\partial q_j} \frac{d q_j}{dt}
\ee

We now define a counterdiabatic Hamiltonian
\be
\label{eq:qh1cd3d} 
\hat H_{CD}(t) = \frac{\hat {\bf p} \cdot \hat {\bf v} + \hat {\bf v} \cdot \hat {\bf p}}{2} \quad,\quad \hat {\bf v}(t) = {\bf v}(\hat {\bf q},t)
\ee
Using $\dot\phi + (1/2)(\nabla\cdot{\bf v})\phi + {\bf v}\cdot \nabla\phi = 0$ (which follows from Eq.~\ref{eq:continuity3d}), it is readily verified that the wavefunction
\be
{\psi}_{ad}({\bf q},t) = \phi({\bf q},t)\, e^{i\alpha(t)}
\quad,\quad
\alpha(t)= - \frac{1}{\hbar} \int_0^t E(t^\prime)\, dt^\prime
\ee
is a solution of the Schr\" odinger equation $i\hbar\, \partial_t{\psi}_{ad} = ( \hat H_0+\hat H_{CD} ) {\psi}_{ad}$.

Now let us suppose that the field ${\bf v}({\bf q},t)$ can be chosen to be curl-free:
\be
\label{eq:curl-free-v}
\nabla\times{\bf v} = {\bf 0}
\ee
We can then introduce a function $S({\bf q},t)$ that satisfies
\be
\label{eq:S3d}
\nabla S = m {\bf v}
\ee
which allows us to rewrite Eq.~\ref{eq:ai3d} as
\be
a_i({\bf q},t)
= \frac{1}{m} \frac{\partial}{\partial q_i} \left[ \frac{\partial S}{\partial t} + \frac{ (\nabla S)^2}{2m} \right]
\label{eq:3d_adef}
\ee
We have used both Eqs.~\ref{eq:curl-free-v} and \ref{eq:S3d} in going from Eq.~\ref{eq:ai3d} to Eq.~\ref{eq:3d_adef}.
If we now define $U_{FF}({\bf q},t)$ by the equation
\be
\partial_t S  + \frac{(\nabla S)^2}{2m} + U_{FF} = 0
\ee
then Eq.~\ref{eq:3d_adef} implies $-\nabla U_{FF} = m{\bf a}$ (compare with Eq.~\ref{eq:qu1ff}).
It is now a matter of algebra to verify that $\bar\psi \equiv \phi e^{i\alpha} e^{iS/\hbar}$
obeys the Schr\" odinger equation $i\hbar\, \partial_t\bar\psi = ( \hat H_0+\hat U_{FF} ) \bar\psi$.
Since $S({\bf q},t)$ is a constant outside the interval $0<t<\tau$, we see that the addition of the fast-forward potential $U_{FF}$ causes the chosen eigenstate of the initial Hamiltonian to evolve to the corresponding eigenstate of the final Hamiltonian.

As in the one-dimensional case, divergences in ${\bf v}({\bf q},t)$ may arise whenever $\phi({\bf q},t) = 0$, potentially causing the method to break down for excited eigenstates.
This issue deserves further exploration.

%%%%%%%%%%%%%%%%%%%%%%%%%%%%%%%%%%%%%%%%%%%%%%%%%%%%%%%%%%%%%%%%%%%%%%%%%%%%%%%%%%%%%%%%%%%%%%%%%%%%%%%%%%%%%%%%%%%%%%%%%%%%%%%%%%%%%%%%%%%%%%%%%%%%%%%%%%%%%%%%%%%%
\section{Classical shortcuts}
\label{sec:classical}

In this section we briefly review the procedure developed in Ref.~\cite{Jarzynski17} for obtaining counterdiabatic and fast-forward shortcuts for a classical Hamiltonian of interest
\begin{equation}
\label{eq:HofI-cl}
H_0(q,p,t) = \frac{p^2}{2m} + U_0(q,t)
\end{equation}
As in Sec.~\ref{sec:quantum}, $H_0$ varies with time only during the interval $0 \le t \le \tau$, and this time-dependence is turned on and off smoothly, as per comments following Eq.~\ref{eq:HofI}.
Let $I(q,p,t) = \oint p^\prime\, dq^\prime$ denote the classical action, where the integral is taken around an {\it energy shell}, that is a level curve of $H_0$, containing the point $(q,p)$.
The action is equal to the volume of phase space enclosed by the energy shell, and it is an adiabatic invariant~\cite{Goldstein80}: in the limit of adiabatic (quasi-static) driving, the value of $I(q,p,t)$ remains constant along a trajectory evolving under $H_0(q,p,t)$.
Thus, the {\it adiabatic evolution} is identified in the classical case with the preservation of the action $I$ along the trajectory.

If $I_0$ is the constant value of the action along a trajectory evolving under adiabatic conditions, then the value of the Hamiltonian along the trajectory, $\bar E(t) \equiv H_0(q(t),p(t),t)$, is determined by the condition
\be
\label{eq:adiabaticEnergy}
\oint_{\bar E(t)} p^\prime\, dq^\prime = I_0
\ee
The notation indicates that the line integral is carried out around the energy shell $H_0(q,p,t) = \bar E(t)$.
We will refer to this shell as the {\it adiabatic energy shell}, and to $\bar E(t)$ as the {\it adiabatic energy}.
The action value $I_0$ and the adiabatic energy $\bar E(t)$ are classical analogues of the quantum number $n$ and eigenenergy $E_n(t)$.

We now assume that $H_0$ varies at an arbitrary -- i.e.\ non-adiabatic -- rate, but we continue to use the term adiabatic energy to refer to $\bar E(t)$ defined by Eq.~\ref{eq:adiabaticEnergy}, for chosen value of action, $I_0$.
For a trajectory with initial action $I_0$, we wish to construct a counterdiabatic Hamiltonian $H_{CD}(q,p,t)$ and a fast-forward potential $U_{FF}(q,t)$ such that:
(1) if the trajectory evolves under $H_0 + H_{CD}$, it remains on the adiabatic energy shell at all times, that is, $I(t) = I_0$; and
(2) if the trajectory evolves under $H_0 + U_{FF}$, it returns to the adiabatic energy shell at the final time: $I(\tau) = I(0) = I_0$.
Here and below, $I(t) = I(q(t),p(t),t)$ denotes the value of the action function along the trajectory.

\begin{figure} 
\centering
\includegraphics[width=0.45\textwidth]{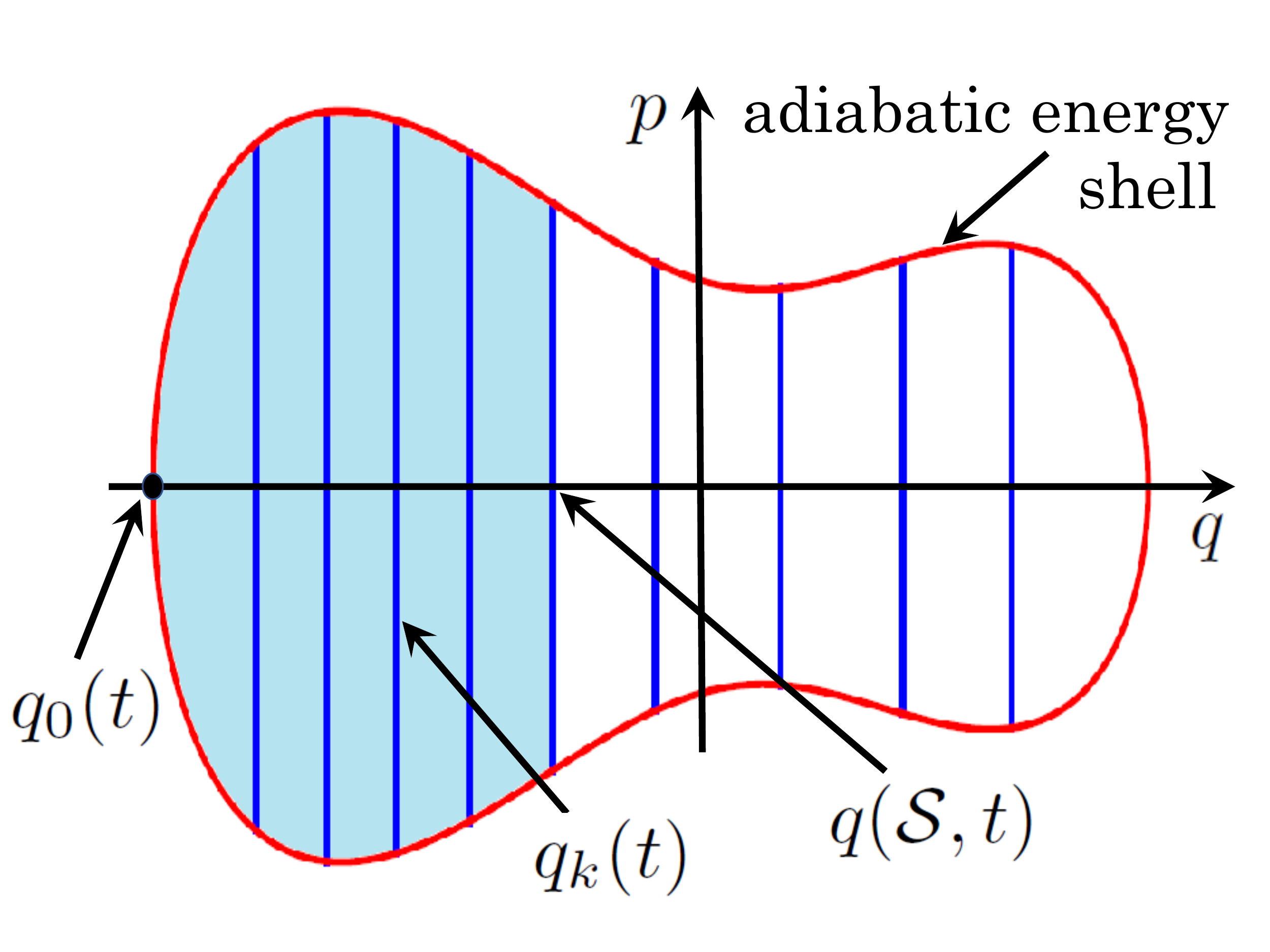}
\caption{\label{Fig:EnergyShell} %(Color online)
The closed red curve, with upper and lower branches $\pm\bar p(q,t)$ (Eq.~\ref{eq:pbar}), depicts the adiabatic energy shell $\bar E(t)$ in phase space. The blue vertical lines divide $\bar E(t)$ into $K\gg 1$ strips of equal phase space volume.
$q(\mathcal{S},t)$ is the right boundary of the shaded region, of phase space volume $\mathcal{S}$.
The parametric motion of the vertical lines defines the velocity and acceleration fields $v(q,t)$ and $a(q,t)$.}
\end{figure} 

To construct these shortcuts for a given choice of $I_0$, let
\be
\label{eq:pbar}
\bar p(q,t) = [ 2m (\bar E - U_0) ]^{1/2}
\ee
specify the upper branch of the adiabatic energy shell, and let
\be
\mathcal{S}(q,t)= 2 \int_{q_0(t)}^q \bar p(q^\prime,t) \cdot dq^\prime
\label{eq:Int_action}
\ee
denote the volume of phase space enclosed by the adiabatic energy shell $\bar E(t)$ between the left turning point $q_0(t)$ and a point $q$. The function $\mathcal{S}(q,t)$ is monotonically increasing and therefore can be inverted to obtain $q(\mathcal{S},t)$, which in turn is used to define flow fields
\begin{subequations}
\label{eq:classicalFlowFields}
\ba
\label{eq:cvdef}
v(q,t) &=& \frac{\partial}{\partial t} q(\mathcal{S},t) =  -\frac{\partial_t \mathcal{S}}{\partial_q \mathcal{S}} \\
\label{eq:cadef}
a(q,t) &=& \frac{\partial^2}{\partial t^2} q(\mathcal{S},t) = v^\prime v + \dot v 
\ea
\end{subequations}
These flow fields are pictured by dividing the adiabatic energy shell into $K\gg 1$ strips enclosing equal phase space volume, delimited by lines drawn at locations $\{ q_k(t) \}$; see Fig.~\ref{Fig:EnergyShell}.
The fields $v$ and $a$ describe the motion of these lines as the parameter $t$ is varied: $\dot q_k = v(q_k,t)$ and $\ddot q_k = a(q_k,t)$.

Using these flow fields we now define a counterdiabatic Hamiltonian
\begin{subequations}
\label{eq:ch1}
\be
\label{eq:ch1cd}
H_{CD}(q,p,t) = pv(q,t)
\ee
and a fast-forward potential $U_{FF}$ that satisfies
\be
\label{eq:cu1ff}
-\partial_q U_{FF}(q,t)= ma(q,t)
\ee
\end{subequations}
both of which vanish for $t\notin(0,\tau)$.
As shown in Ref.~\cite{Jarzynski17}, if a trajectory evolves under the Hamiltonian $H_0+H_{CD}$, then the action is preserved along the entire trajectory: $I(t) = I_0$ for all $t$.
If the trajectory instead evolves under $H_0+U_{FF}$, then the initial and final action values are equal, $I(0) = I(\tau) = I_0$,
though $I(t)\ne I_0$ at intermediate times.
In both cases the action $I(q,p,t)$ is defined with respect to the Hamiltonian $H_0(q,p,t)$, and the construction of $H_{CD}$ and $U_{FF}$ depends on the choice of action $I_0$.

\begin{figure}[tbp]
   \subfigure{
   \label{fig:initialConditions}
   \includegraphics[trim = 2in 5in 0in 0in, scale=0.30]{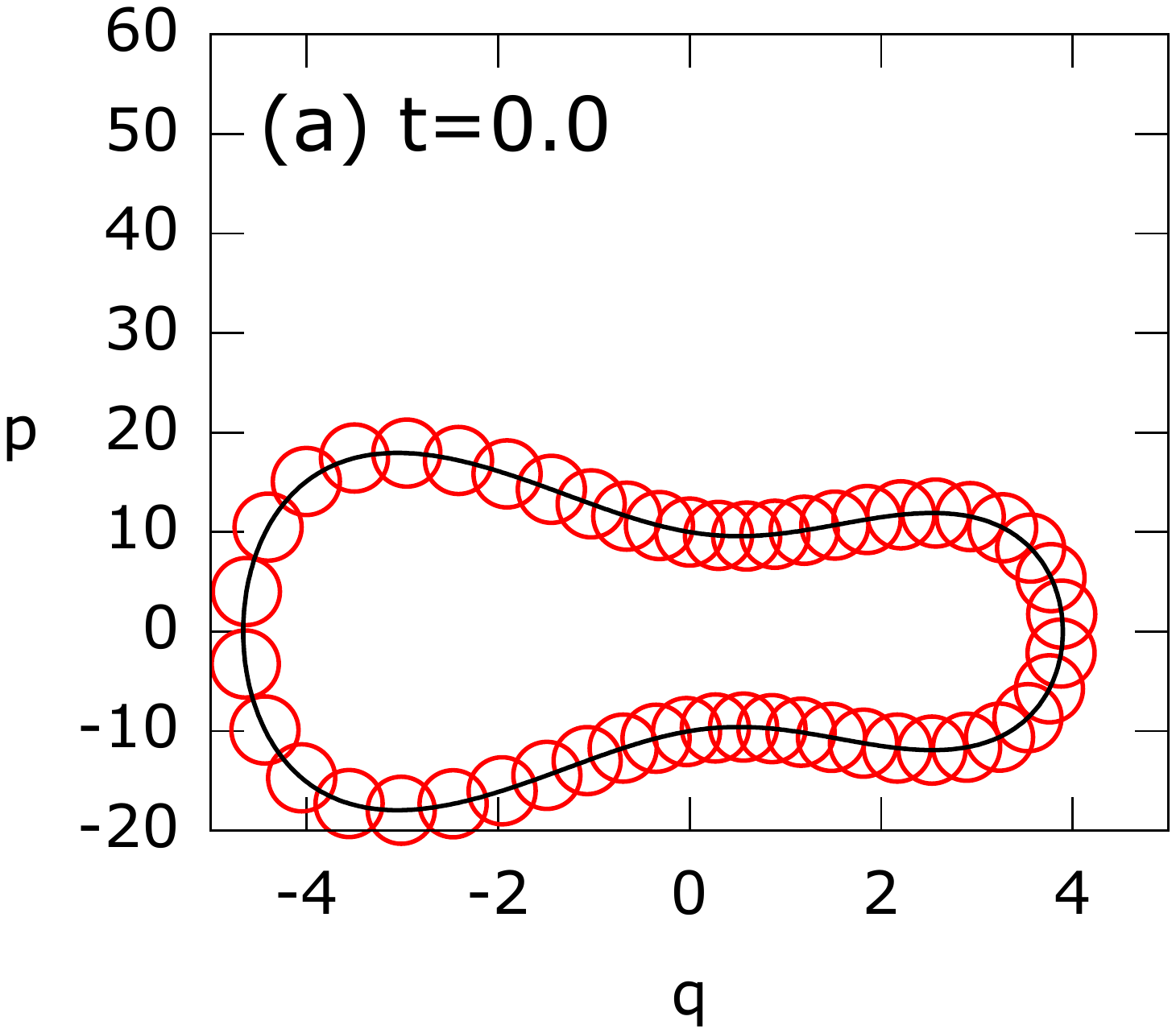}
   }
   \subfigure{
   \label{fig:intermediateConditions_FF}
   \includegraphics[trim = 2in 5in 0in 0in, scale=0.30]{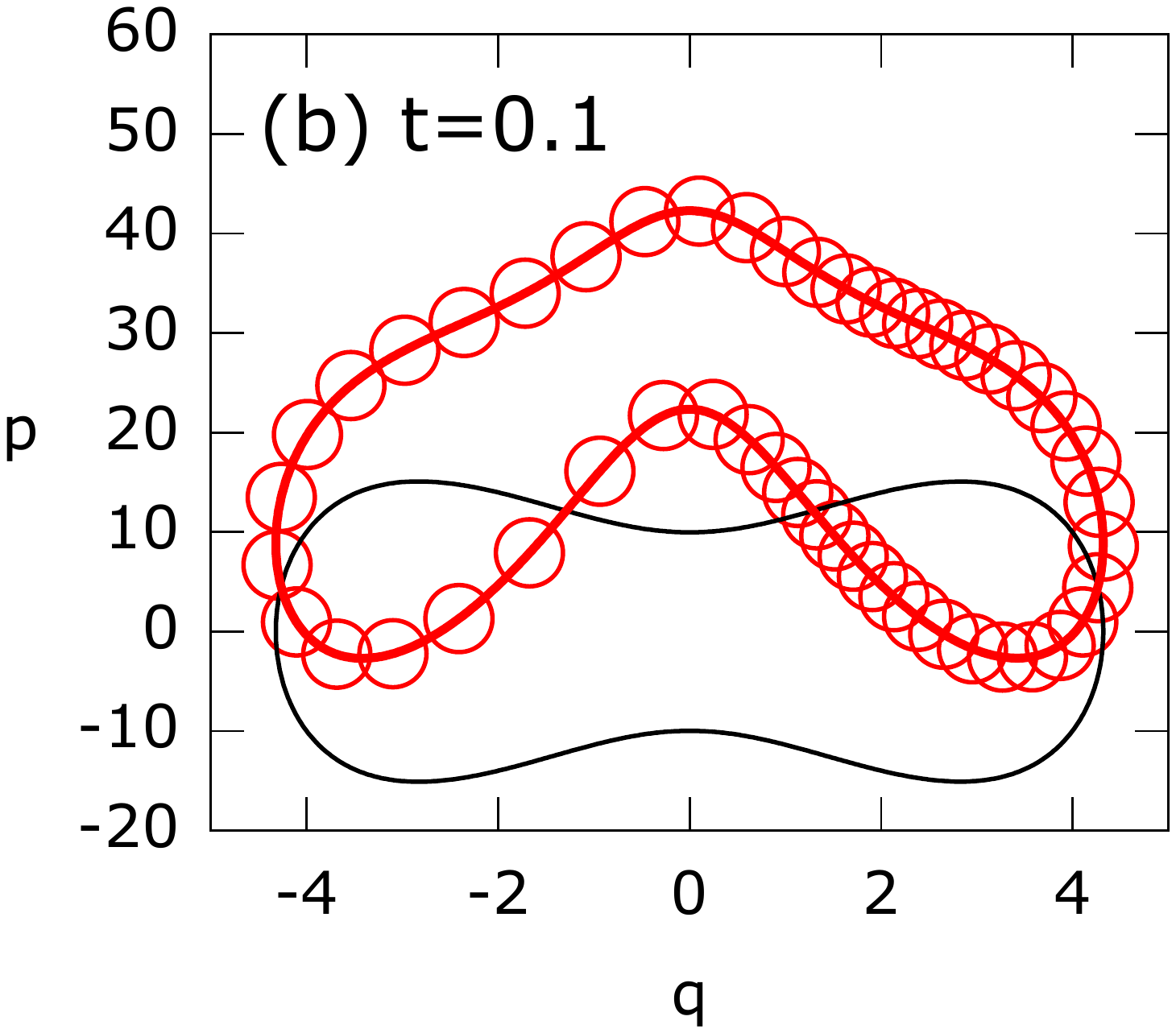}
   }
   \subfigure{
   \label{fig:finalConditions_FF}
   \includegraphics[trim = 2in 5in 0in 0in, scale=0.30]{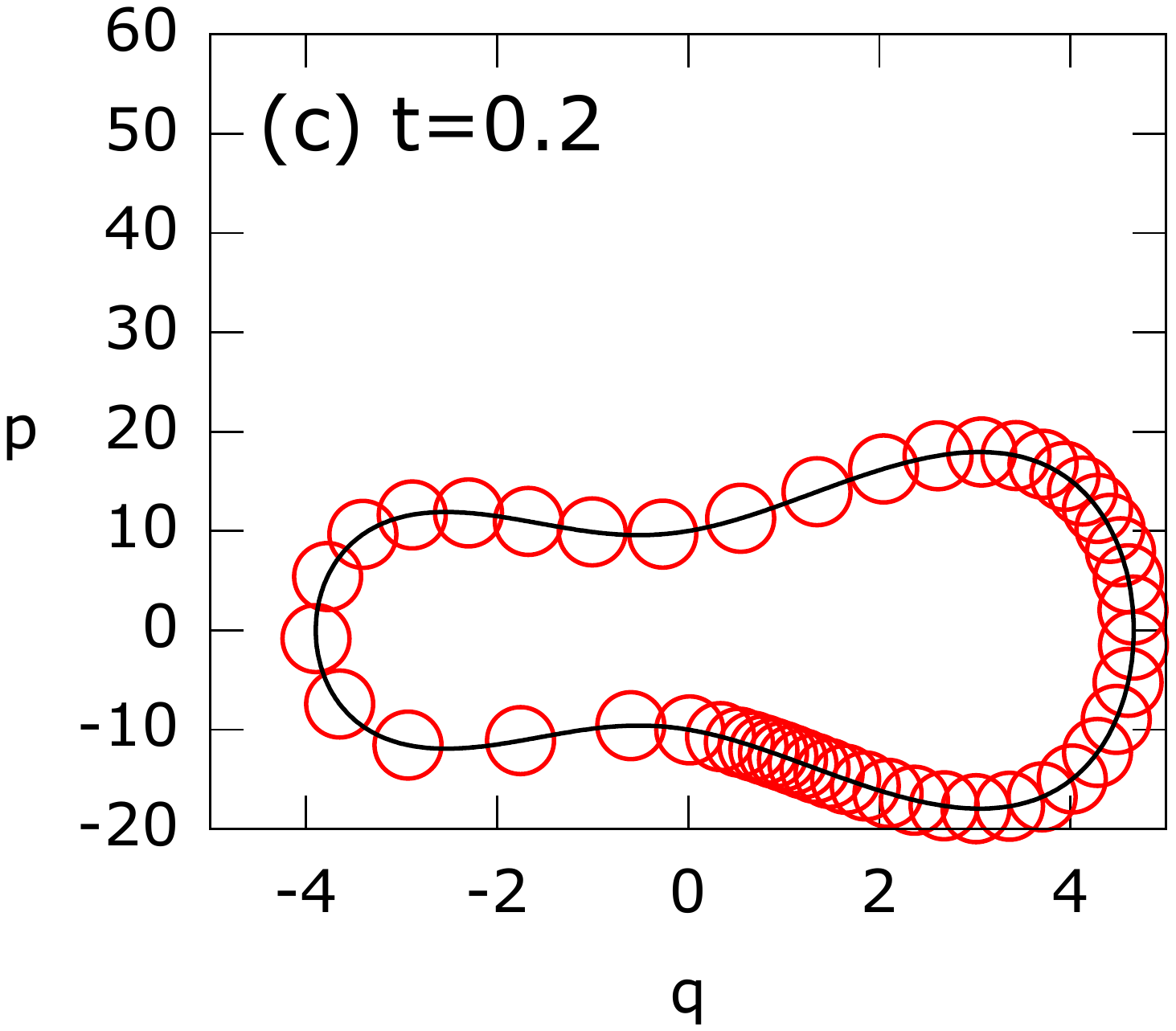}
   }
\caption{ %(Color online)
The red circles show fifty trajectories evolving in phase space under the Hamiltonian $H_0 + U_{FF}$,
where $H_0(q,p,t)$ is given by Eq.(22) of Ref.~\cite{Jarzynski17},
and $U_{FF}(q,t)$ was constructed using Eq.~\ref{eq:cu1ff} of the present paper.
The duration of the process is $\tau=0.2$.
The thin black curve is the adiabatic energy shell.
In frame (b), the thick red curve is obtained by displacing, or ``boosting'', every point on the energy shell by an amount $mv(q,t)$, along the momentum direction.
}
\label{fig:proofOfPrinciple}
\end{figure}

Figure~\ref{fig:proofOfPrinciple} illustrates evolution under the fast-forward Hamiltonian $H_0+U_{FF}$.
The trajectories begin on the adiabatic energy shell, they evolve away from it at intermediate times, and return to it at the final time.

\section{Comparison of quantum and classical shortcuts}
\label{sec:comparison}

As proposed in Ref.~\cite{Jarzynski13} and illustrated in Ref.~\cite{Patra16}, a classical counterdiabatic Hamiltonian emerges when the right side of Eq.~\ref{eq:HCD} is evaluated in the semiclassical limit.
Similarly, it is natural to speculate that the classical shortcuts of Sec.~\ref{sec:classical} are the semiclassical limit of the quantum shortcuts of Sec.~\ref{sec:quantum}.
In that case the close similarity between Eqs.~\ref{eq:ch1cd}, \ref{eq:cu1ff} and Eqs.~\ref{eq:qh1cd}, \ref{eq:qu1ff} would simply reflect the Correspondence Principle.
In the following paragraph we address this issue by asking whether the flow fields $v(q,t)$ and $a(q,t)$ defined in Sec.~\ref{sec:classical} emerge from those of Sec.~\ref{sec:quantum} in the semiclassical limit ($\hbar\rightarrow 0$).
We will temporarily use the superscript $Q$ (for ``quantum'') to denote certain quantities defined in Sec.~\ref{sec:quantum}, $SC$ to denote their semiclassical limits, and $C$ to denote quantities defined in Sec.~\ref{sec:classical}.

When we consider the semiclassical limit of the field $v^Q(q,t)$ (Eq.\ref{vel}), we immediately run into a difficulty: the divergences discussed in Sec.~\ref{subsec:divergences} proliferate in this limit, as the number of nodes of $\phi$ becomes large.
This proliferation of divergences (nodes) arises from the rapid spatial oscillations of high-lying eigenstates $\phi(q,t)$.
To obtain a non-singular velocity field, we replace the oscillatory probability density $\phi^2$ (used to construct $v^Q$) by a locally averaged counterpart, $\overline{\phi^2}$, that smooths over these oscillations.
The semiclassical limit of $\overline{\phi^2}$ is the microcanonical probability distribution, projected from phase space onto the coordinate axis~\cite{Griffiths04}:
\be
\lim_{\hbar\rightarrow 0} \overline{\phi^2}(q,t) = \mu(q,t) \propto \int dp \, \delta ( \bar E - H_0 ) \propto \frac{1}{\bar p(q,t)}
\ee
with $\bar p$ given by Eq.~\ref{eq:pbar}.
Using $\mu$ in place of $\phi^2$ in Eq.~\ref{I} we obtain
\be
\mathcal{I}^{SC}(q,t) = \int_{- \infty}^q  \mu(q^\prime,t) \, dq^\prime.
\ee
We use this function to define $v^{SC} = -\partial_t \mathcal{I}^{SC}/\partial_q \mathcal{I}^{SC}$,
which is free of divergences and can be viewed as the semiclassical limit of $v^Q = -\partial_t \mathcal{I}^{Q}/\partial_q \mathcal{I}^{Q}$ (Eq.~\ref{vel}).
Comparing $v^{SC}(q,t)$ with the field $v^C(q,t)$ defined by Eq.~\ref{eq:cvdef}, we see that while one is constructed from the integrated microcanonical distribution $\mathcal{I}^{SC} = \int^q \mu\, dq^\prime$, the other  is constructed in terms of the phase space enclosed by the energy shell, $\mathcal{S} = \int^q \bar p\, dq^\prime$.
Therefore, in general, the two fields differ: $v^{SC} \ne v^{C}$.
We conclude that Eq.~\ref{eq:ch1cd} should not be viewed as the semiclassical limit of Eq.~\ref{eq:qh1cd}.
Similar comments apply to the acceleration field $a(q,t)$.

We summarize the situation as follows: while the flow fields $v$ and $a$ are defined similarly in the quantum and classical cases (compare Figs.~\ref{Fig:I_lines} and \ref{Fig:EnergyShell}), and while the construction of counterdiabatic and fast-forward terms from the flow fields is essentially identical in the two cases, the Correspondence Principle does not provide an adequate explanation for this striking similarity.
We also note that scale-invariant driving (Eq.~\ref{SI_H0}) provides an exception to this general conclusion: in that case the quantum and classical flow fields are in fact identical~\cite{Deffner14,Jarzynski17}.

As a final item of semiclassical comparison, let us consider trajectories evolving under the classical Hamiltonian $H_0+U_{FF}$, with initial conditions sampled from the adiabatic energy shell (Fig.~\ref{fig:proofOfPrinciple}).
It was shown in Ref.~\cite{Jarzynski17} that the function
\be
J(q,p,t) = I(q,p-mv(q,t),t)
\ee
remains constant along these trajectories: $J(t) = I_0$ for all $t$.
This is illustrated in Fig.~\ref{fig:intermediateConditions_FF}, where the thick red curve is obtained by ``boosting'' the thin black curve -- the adiabatic energy shell -- by an amount $mv(q,t)$ along the momentum direction.
Now consider the fast-forward wavefunction $\bar\psi=  \phi \, e^{i\alpha}\, e^{iS/\hbar}$ (Eq.~\ref{eq:ansatz}) evolving under $\hat H_0+\hat U_{FF}$
Let us approximate the eigenstate $\phi(q,t)$ by the WKB form~\cite{Griffiths04}
\be
\label{eq:wkb}
\phi = A_+ e^{+(i/\hbar) \int^q \bar p \, dq^\prime} + A_- e^{-(i/\hbar) \int^q \bar p \, dq^\prime}
\ee
where $\vert A_\pm(q,t)\vert \propto \sqrt{1/\bar p}$.
The terms on the right side of Eq.~\ref{eq:wkb} represent a right-moving wave train and a left-moving wave train, with local momenta corresponding to the upper and lower branches $\pm\bar p$ of the adiabatic energy shell (Fig.~\ref{Fig:EnergyShell}).
Then for the fast-forward wavefunction we get
\be
\label{eq:boostedpsi}
\bar\psi = \phi\, e^{i\alpha}\, e^{iS/\hbar}
= A_+ e^{i\alpha} \, e^{(i/\hbar) \int^q (\bar p+mv) \, dq^\prime} + A_- e^{i\alpha} \, e^{(i/\hbar) \int^q (-\bar p+mv) \, dq^\prime}
\ee
since $S = \int^q mv \, dq^\prime$ (Eq.~\ref{eq:S}).
The terms in Eq.~\ref{eq:boostedpsi} are wave trains with local momenta $\pm\bar p + mv$.
Thus the fast-forward wavefunction $\bar\psi$ is represented, in the WKB sense, by a ``boosted'' adiabatic energy shell similar to the one shown as a thick red curve in Fig.~\ref{fig:intermediateConditions_FF}.
Although this interpretation provides a neat correspondence between the quantum and classical fast-forward methods, it should not be taken too literally, since the fast-forward method of Sec.~\ref{sec:quantum} generally applies only to the ground state (as discussed earlier), where the WKB approximation (Eq.~\ref{eq:wkb}) is not generally accurate.

%%%%%%%%%%%%%%%%%%%%%%%%%%%%%%%%%%%%%%%%%%%%%%%%%%%%%%%%%%%%%%%%%%%%%%%%%%%%%%%%%%%%%%%%%%%%%%%%%%%%%%%%%%%%%%%%%%%%%%
\section{Shortcuts for stochastic systems}
\label{sec:stochastic}

\begin{figure} 
\centering
\includegraphics[width=0.45\textwidth]{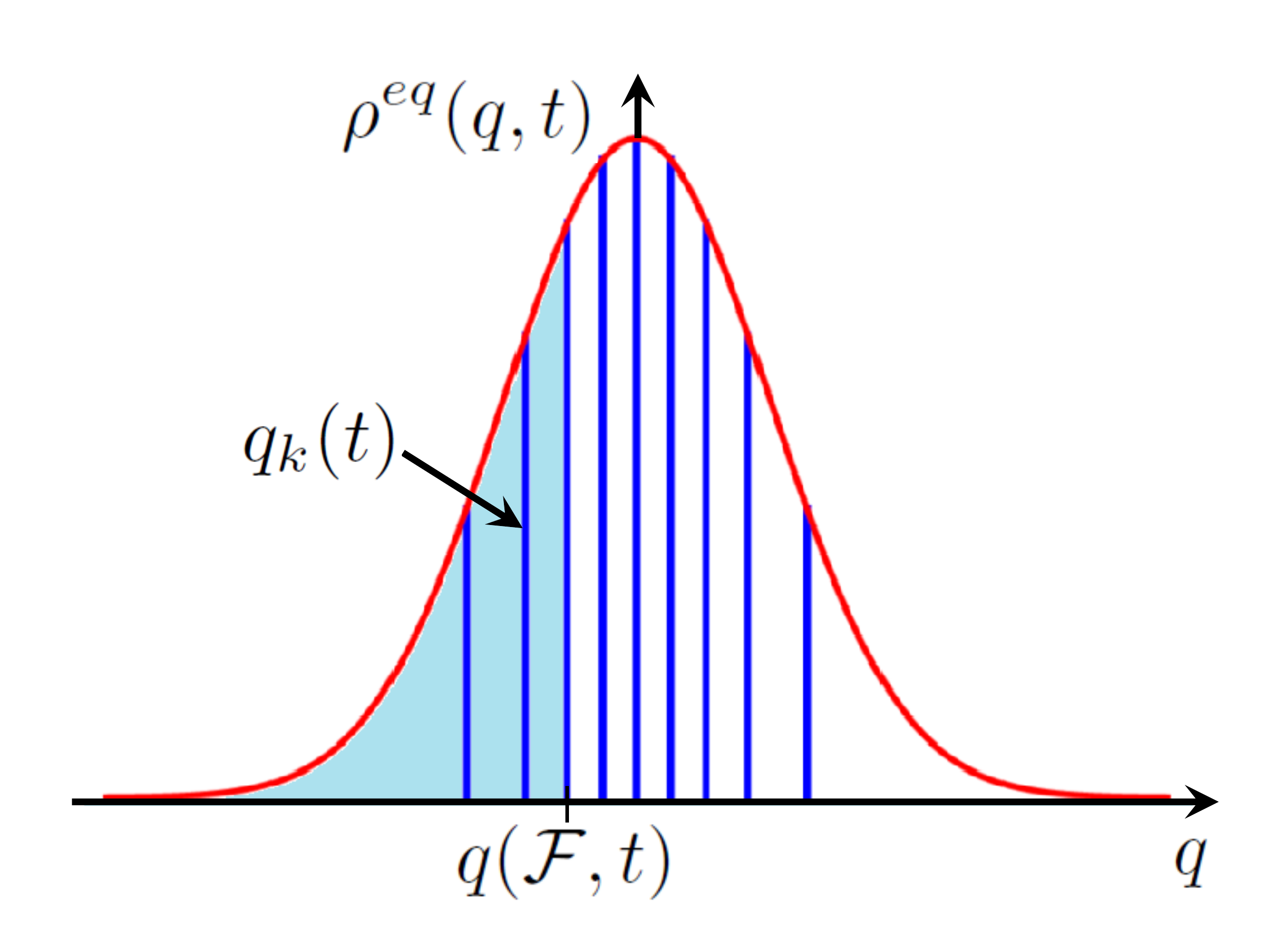}
\caption{\label{fig:Prob_lines} %(Color online)
The blue lines divide the equilibrium distribution into strips of equal area. 
$q(\mathcal{F},t)$ is the right boundary of the shaded region, which has area $\mathcal{F}$.
The velocity field $v(q,t)$ describes the motion of the vertical lines with $t$ (Eq.~\ref{flowf_stoc}).
}
\end{figure} 

Finally, let us consider an overdamped Brownian particle in a time-dependent potential $U_0(q,t)$, which varies smoothly for $0\le t\le\tau$ but is fixed outside this interval.
The particle is in contact with a thermal reservoir at temperature $T$.
The interactions with the degrees of freedom of the reservoir give rise to the random and dissipative forces that characterize Brownian dynamics.

We will work in the ensemble picture, in which the dynamics are described by a probability density function $\rho(q,t)$ that evolves according to the Fokker-Planck equation
\ba
\partial_t \rho =  \frac{1}{\gamma} \partial_q \left[ (\partial_q U_0) \rho \right] + D \partial_q^2 \rho \equiv  \hat{\mathcal{L}}_0(t) \rho,
 \label{FP}
\ea
The friction and diffusion coefficients, $\gamma$ and $D$, obey the Einstein-Smoluchowski relation, $\gamma D = k_BT\equiv 1/\beta$, where $k_B$ is the Boltzmann constant.
The equilibrium distribution associated with the potential $U_0(q,t)$ is given by
\be
\rho^{eq}(q,t) = \frac{1}{Z(t)} \exp\left[-\beta U_0(q,t)\right]
\label{rhoeq}
\ee
It is straightforward to verify the identity
\ba
\hat{\mathcal{L}}_0(t) \rho^{eq}(q,t) =0
\label{eqbmFP}
\ea
which confirms that the equilibrium distribution is a stationary solution of the dynamics, when the potential does not vary with time.
When $U_0(q,t)$ changes quasi-statically, then the slowly varying $\rho^{eq}(q,t)$ is a solution of Eq.~\ref{FP} (both sides tend toward zero in that limit), as expected for a reversible process.
Therefore, the {\it adiabatic evolution} is identified by the ensemble evolving through the continuous sequence of equilibrium states $\rho^{eq}(q,t)$.~\footnote{
Here we use the term {\it adiabatic} consistently with its usage in the rest of the paper, namely to denote a slow process.
This differs from its usage in thermodynamics, where an adiabatic process is one in which heat is not absorbed or released by a system.
The dual meaning of this term is unfortunate.
}

We now consider the case in which the potential $U_0(q,t)$ is varied at an arbitrary rate.
To this potential we will add a counterdiabatic term $U_{CD}(q,t)$, so that the system evolves under
\ba
\partial_t \rho &=& \frac{1}{\gamma} \partial_q \left\{ \left[ \partial_q (U_0 + U_{CD}) \right] \rho \right\} + D \partial_q^2 \rho  \non \\
&=& \hat{\mathcal{L}}_0 \rho + \frac{1}{\gamma} \partial_q \left[ (\partial_q U_{CD}) \rho \right]
 \label{FPfull}
\ea
We wish to design $U_{CD}(q,t)$ so as to achieve adiabatic evolution, i.e.\ so that $\rho^{eq}(q,t)$ (Eq.~\ref{rhoeq}) is an exact solution of Eq.~\ref{FPfull}.
This problem has recently been studied by Li, Quan and Tu~\cite{Li17}, who use a somewhat different derivation to arrive at the same solution as the one we obtain below (Eq.~\ref{Ucd}).

We define an integrated distribution
\ba
\mathcal{F}(q,t) = \int_{- \infty}^{q} \rho^{eq}(q',t) dq'
\label{F} 
\ea
similar to $\mathcal{I}(q,t)$ (Eq.~\ref{I}).
Inverting this function to obtain $q(\mathcal{F},t)$ (see Fig.~\ref{fig:Prob_lines}), we construct a velocity flow field
\ba
v(q,t)=\frac{\partial}{\partial t} q(\mathcal{F},t) = -\frac{\partial_t \mathcal{F}}{\partial_q \mathcal{F}}.
\label{flowf_stoc}
\ea
Rearranging this result as $\partial_t \mathcal{F}+v \partial_q \mathcal{F}=0$ and differentiating with respect to $q$ produces the continuity equation
\ba
\partial_t \rho^{eq} + \partial_q(v \rho^{eq})=0.
\label{cont_stoc}
\ea

Since we wish $\rho^{eq}(q,t)$ to be a solution of Eq.~\ref{FPfull}, we use Eqs.~\ref{eqbmFP} and \ref{cont_stoc} to rewrite that equation as follows:
\be
-\partial_q (v \rho^{eq}) = \frac{1}{\gamma} \partial_q \left[ (\partial_q U_{CD}) \rho^{eq} \right]
\ee
Integrating both sides gives
\be
-v \rho^{eq} - J(t) = \frac{1}{\gamma} (\partial_q U_{CD}) \rho^{eq}
\ee
Here, $J(t)$ is an arbitrary function of time that we set to zero, for convenience, arriving at the result
\be
\label{Ucd}
-\partial_q U_{CD}(q,t) = \gamma v(q,t)
\ee

Eq.~\ref{Ucd} defines $U_{CD}(q,t)$ up to an additive function of time that has no influence on the dynamics, and which can be adjusted so that $U_{CD} = 0$ for $t\notin(0,\tau)$.
The potential $U_{CD}(q,t)$ has the desired counterdiabatic property: when the system evolves in the time-dependent potential $U_0+U_{CD}$, it remains in equilibrium (with respect to $U_0)$ over the entire duration of the process.
Our potential $U_{CD}$ is equivalent to the auxiliary potential obtained by Li {\it et al}~\cite{Li17}, as can be seen by differentiating both sides of Eq.\ 12 of Ref.~\cite{Li17} with respect to $x$.

Eq.~\ref{Ucd} has elements in common with both the counterdiabatic and fast-forward shortcuts of previous sections.
It is counterdiabatic in that the system follows the adiabatic evolution (it remains in the state $\rho^{eq}$) at all times.
Moreover, $U_{CD}$ is given in terms of the velocity field $v$ (compare Eqs.~\ref{eq:qh1cd}, \ref{eq:ch1cd}, and \ref{Ucd}), rather than the acceleration field $a$.
However, just as with the fast-forward shortcuts described earlier, $U_{CD}$ is {\it local}, i.e.\ it is a time-dependent potential (compare Eqs.~\ref{eq:qu1ff}, \ref{eq:cu1ff}, and \ref{Ucd}).
Also, as in Secs.~\ref{sec:quantum} and \ref{sec:classical}, Eq.~\ref{Ucd} defines the auxiliary potential only up to an arbitrary function of time that does not affect the dynamics.

When our Brownian particle evolves under $U_0(q,t)$ alone, the state of the ensemble, $\rho(q,t)$, {\it lags} behind the instantaneous equilibrium state, $\rho^{eq}(q,t)$, as illustrated in Ref.~\cite{Vaikuntanathan09}, Figure 1.
The addition of the counterdiabatic potential $U_{CD}(q,t)$ eliminates this lag.
Lagging distributions are relevant for numerical free-energy estimation methods, where the lag gives rise to poor convergence of the free-energy estimate.
Vaikuntanathan and Jarzynski~\cite{Vaikuntanathan08} have developed a method in which this lag is reduced or eliminated by the addition of an artificial flow field to the dynamics, although in Ref.~\cite{Vaikuntanathan08} this flow field was not related to an auxiliary potential $U_{CD}$.
Comparing Eq.~\ref{cont_stoc} above with Eq.~15 of Ref.~\cite{Vaikuntanathan08}, we see that our field $v$ is equivalent to the {\it perfect flow field} ($u^*$) that ``escorts'' the system faithfully along the equilibrium path.

As a simple example, which makes a connection to the {\it engineered swift equilibration} approach recently developed and validated experimentally by Mart\' inez {\it et al}~\cite{Martinez16}, we consider a time-dependent harmonic potential,
\be
U_0(q,t) = \frac{1}{2} \kappa_0(t)q^2
\label{V0}
\ee
The instantaneous equilibrium distribution is
\be
\rho^{eq}(q,t)=\sqrt{\frac{\sigma}{\pi}} \exp(-\sigma q^2) \quad,\quad \sigma(t)\equiv\beta\kappa_0(t)/2 .
\label{rhoeqsho}
\ee
Eq.~\ref{F} then gives
\be
\mathcal{F}(q,t) = \frac{1}{2} \left[ 1+\textrm{erf}(\sqrt{\sigma(t)}q) \right],
\label{Fsho}
\ee
where
$\textrm{erf}(\cdot)$ is the Gaussian error function.
In turn, Eq.~\ref{flowf_stoc} yields $v(q,t)=-\dot\sigma/2\sigma$,
and from there we use Eq.~\ref{Ucd} to obtain
\be
U_{CD}(q,t)= \left(\frac{\gamma \dot{\sigma}}{2 \sigma}\right) \frac{q^2}{2}.
\ee
Therefore, under a harmonic trap of stiffness 
\be
\kappa(t) = \kappa_0+ \frac{\gamma \dot{\sigma}}{2 \sigma} = \kappa_0+ \frac{\gamma \dot{\kappa_0}}{2 \kappa_0}
\label{compk}
\ee
the ensemble remains in the equilibrium state $\rho^{eq}$ (Eq.~\ref{rhoeqsho}) at all times.

Rearranging Eq.~\ref{compk} and using $\sigma=\beta\kappa_0/2$, we get
\be
\label{eq:swift}
\frac{\dot{\sigma}}{\sigma} = \frac{2 \kappa}{\gamma} -  \frac{4k_B T \sigma}{\gamma}.
\ee
This result is identical to Eq.~6 of Mart\' inez {\it et al}~\cite{Martinez16}, where the goal was to bring the system rapidly to the final equilibrium state, without concern for the intermediate states visited along the way.
Our approach achieves the same result by guiding the system along the equilibrium path during the entire process.
Eq.~\ref{eq:swift} was also obtained by Schmiedl and Seifert, in the contexts of optimal finite-time control~\cite{Schmiedl07} and stochastic heat engines~\cite{Schmiedl08}.

While our analysis has been restricted to overdamped Brownian motion, Le Cununder and colleagues~\cite{LeCununder16} have recently used a micromechanical cantilever to implement engineered swift equilibration for an underdamped harmonic oscillator.
For underdamped motion in a general one-dimensional, time-dependent potential, Li, Quan and Tu~\cite{Li17} have proposed a momentum-dependent counterdiabatic term that achieves the desired adiabatic evolution.
It remains to be seen whether this progress will lead to expressions for a momentum-{\it independent} counterdiabatic potential that extends the fast-forward method to underdamped Brownian dynamics beyond the harmonic regime.

\section{Conclusions}
\label{sec:conclusion}

We have developed a framework for constructing counterdiabatic and fast-forward shortcuts for quantum, classical and stochastic systems. This framework is organized around velocity and acceleration flow fields $v(q,t)$ and $a(q,t)$, which describe the time-dependence of the desired adiabatic evolution. Once the flow fields have been determined, the shortcuts are given by simple expressions involving these fields (Eqs.~\ref{eq:qh1cd}, \ref{eq:qu1ff}, \ref{eq:ch1cd}, \ref{eq:cu1ff}, \ref{Ucd}).

The flow fields themselves are defined similarly, but not identically, in the three cases -- quantum, classical, and stochastic. In each case these fields can be pictured in terms of the evolution of an appropriately defined ``picket fence'' of lines (Figs.~\ref{Fig:I_lines}, \ref{Fig:EnergyShell}, \ref{fig:Prob_lines}) that glide around as time is varied parametrically. Formally, the fields $v$ and $a$ are constructed from integrated functions, $\mathcal{I}$, $\mathcal{S}$ and $\mathcal{F}$, that define the picket fence.

Since both the classical and quantum cases involve an isolated system evolving under a time-dependent Hamiltonian, it might be expected that they are related by the Correspondence Principle. However, we argued in Sec.~\ref{sec:comparison} that in the semiclassical limit, the integrated function used in the quantum case, $\mathcal{I}(q,t)$, does not converge to the one used in the classical case, $\mathcal{S}(q,t)$, even after appropriate local averaging to smooth over divergences. In this sense the quantum shortcuts of Sec.~\ref{sec:quantum} do not converge to the classical ones of Sec.~\ref{sec:classical} as $\hbar\rightarrow 0$. Hence the Correspondence Principle does not provide a satisfactory understanding of the conspicuous similarity between the expressions for $\hat H_{CD}$ and $\hat U_{FF}$ (Eqs.~\ref{eq:qh1cd}, \ref{eq:qu1ff}), and those for $H_{CD}$ and $U_{FF}$ (Eqs.~\ref{eq:ch1cd}, \ref{eq:cu1ff}).

As noted in Sec.~\ref{subsec:divergences}, the nodes of excited energy eigenstates $\phi(q,t)$ generically pose a problem for our method, as they do for the fast-forward approach in general. The divergences in $v(q,t)$ that result from these nodes can be understood intuitively by considering Eq.~\ref{eq:flux}, which gives the probability flux across the $\nu$'th node: $\Phi_\nu = (v-u_\nu) \phi^2$. The two factors on the right represent the flow velocity relative to the motion of the node, $v-u_\nu$, and the local density, $\phi^2$. If we momentarily imagine that $\phi^2$ is very small but non-zero at $q_\nu$, then we see that $v-u_\nu$ must be very large in order to ``push through'' a fixed probability flux -- an apt analogy is water flowing through a pipe that becomes narrow at a certain point. Thus $v-u_\nu$ diverges as $\phi^2\rightarrow 0$: an infinite velocity is required to achieve a finite flux, at vanishing probability density.

When the time-dependence of $\phi^2(q,t)$ is such that there is no flux of probability across nodes, i.e.\ when the probability between neighboring nodes remains constant even as the eigenstate deforms, then the flow fields $v$ and $a$ are non-singular and we expect our method (and more generally the fast-forward approach~\cite{Masuda10}) to work well. This no-flux criterion is satisfied for scale-invariant driving, as well as for the model system studied numerically in Sec.~\ref{subsec:model}. In the latter case the criterion is satisfied because the potential $U_0(q,t)$ (Eq.~\ref{PotRazavy}) is symmetric about the origin. It would be useful to identify a more generic (i.e.\ non-symmetric) potential and eigenstate for which the no-flux criterion is satisfied, and to test whether our method continues to work in that situation. This would provide a more stringent test of the no-flux criterion than the one studied in Sec.~\ref{subsec:model}.

Our framework connects the counterdiabatic and fast-forward approaches. In the quantum case, the fields $v(q,t)$ and $a(q,t)$ provide two mathematical descriptions the same flow of probability. The former defines the counterdiabatic Hamiltonian $\hat H_{CD}$, while the latter (together with the mass, $m$) determines the fast-forward potential $\hat U_{FF}$. It is remarkable that no other input is required to construct these shortcuts. For the moment we lack a deeper or intuitive understanding of why this should be the case. Analogous comments apply to the classical shortcuts of Sec.~\ref{sec:classical}.

Finally, an open problem remains the development of fast-forward shortcuts for excited states of quantum systems, when the no-flux criterion is not satisfied. We are currently investigating whether the insights provided by our approach, based on flow fields, might yield either exact or approximate solutions to this problem. Another question worth exploring is whether our flow-fields method can be extended to obtain shortcuts for open quantum systems~\cite{Jing13,Vacanti14,Song16,Jing16}.

\section*{Acknowledgments}

We gratefully acknowledge financial support from the U.S. National Science Foundation under grant DMR-1506969 (CJ), and the U.S. Army Research Office under contract 
number W911NF-13-1-0390 (AP), and stimulating discussions with Alexander Boyd, Sebastian Deffner, Adolfo del Campo, Anatoli Polkovnikov and Yi\u{g}it Suba\c{s}\i.

\bibliography{References}

\end{document}